\documentclass[12pt]{article}
\pdfoutput=1
\usepackage{makeidx}
\makeindex
\usepackage[a4paper]{geometry}
\usepackage{jheppub,amsmath,amssymb,amsfonts,amsxtra,mathrsfs,graphics,graphicx,amsthm,epsfig,%ytableau,
bm,longtable,float,color,tikz,mathtools,xfrac,footnote,rotating,lscape}
\restylefloat{table}
\usepackage{dsfont}
\usepackage{multicol}
\usepackage{lipsum}
\usepackage{textgreek}
\usetikzlibrary{decorations.pathmorphing}
\usetikzlibrary{decorations.markings}
\usetikzlibrary{quotes,arrows.meta}
\usetikzlibrary{arrows, decorations.markings, calc, fadings, decorations.pathreplacing, patterns, decorations.pathmorphing, positioning}
\usepackage{tikz-cd}

\usepackage{fixmath} % for \mathbold
\usepackage{scalerel}
\newlength\bshft
\def\fakebold#1{\ThisStyle{\ooalign{$\SavedStyle#1$\cr%
  \kern-\bshft$\SavedStyle#1$\cr%
  \kern\bshft$\SavedStyle#1$}}}

\usetikzlibrary{positioning,shapes}
\usetikzlibrary{chains}
\usetikzlibrary{arrows,fit,decorations.pathreplacing}
\tikzstyle{every picture}+=[remember picture]
\tikzstyle{na} = [baseline=-.5ex]

\usepackage{empheq}
\usepackage{multirow}
\usepackage{booktabs}
\usepackage[american]{babel}

\usepackage[utf8]{inputenc}

\usepackage{array,booktabs}

\makeatletter
\newcommand{\vast}{\bBigg@{1}}
\newcommand{\Vast}{\bBigg@{5}}
\makeatother

\usepackage{hyperref}

\numberwithin{equation}{section}

\newcommand{\eg}{\textit{e.g.}}

\newcommand{\ie}{\textit{i.e.}}

\newcommand{\ii}{\mathrm{i}}
\newcommand{\?}{\;\!}

\numberwithin{equation}{section}

\newcommand{\be}{\begin{equation}} \newcommand{\ee}{\end{equation}}
\newcommand{\bea}{\begin{equation} \begin{aligned}} \newcommand{\eea}{\end{aligned} \end{equation}}

\newcommand{\Iprod}[2]{\langle {#1}, {#2} \rangle}

\def\U{\mathrm{U}}
\def\SO{\mathrm{SO}}

\def\SU{\mathrm{SU}}
\def\SL{\mathrm{SL}}

\def\u{\mathsf{u}}
\def\v{\mathsf{v}}
\def\j{\mathsf{j}}
\newcommand{\ex}{{\mathrm{e}}}

\newcommand{\rd}{\mathrm{d}}

\newcommand{\wt}{\widetilde}

\DeclareMathOperator{\tr}{tr}

\newcommand{\cA}{\mathcal{A}}
\newcommand{\cB}{\mathcal{B}}
\newcommand{\cC}{\mathcal{C}}

\newcommand{\cF}{\mathcal{F}}

\newcommand{\cH}{\mathcal{H}}
\newcommand{\cI}{\mathcal{I}}
\newcommand{\cJ}{\mathcal{J}}

\newcommand{\cM}{\mathcal{M}}
\newcommand{\cN}{\mathcal{N}}

\newcommand{\cT}{\mathcal{T}}

\newcommand{\cW}{\mathcal{W}}

\newcommand{\bR}{\mathbb{R}}
\newcommand{\bZ}{\mathbb{Z}}

\newcommand{\fg}{\mathfrak{g}}

\newcommand{\fp}{\mathfrak{p}}

\newcommand{\fR}{\mathfrak{R}}
\newcommand{\fs}{\mathfrak{s}}
\newcommand{\ft}{\mathfrak{t}}

\definecolor{cadmiumgreen}{rgb}{0.0, 0.42, 0.24}

\usepackage[Symbol]{upgreek}
\usepackage{bm}

\usepackage{calligra}
\DeclareMathAlphabet{\mathcalligra}{T1}{calligra}{m}{n}

\theoremstyle{plain}

  \theoremstyle{definition}

\providecommand{\examplename}{Example}
\providecommand{\theoremname}{Theorem}

\makeatletter
\g@addto@macro\bfseries{\boldmath}
\makeatother

\makeatletter
\newcommand*{\rom}[1]{\expandafter\@slowromancap\romannumeral #1@}
\makeatother

\def\CS{\mathcal{CS}}
\def\Nequals#1{$\mathcal{N}=#1$}
\def\pair#1#2{\det(#1,#2)}
\def\ww{} % used to be _w
\title{Anomalies, Black strings \\
and  the charged Cardy formula}

\author[a]{Seyed Morteza Hosseini,}
\author[b]{Kiril Hristov,}
\author[a]{Yuji Tachikawa,}
\author[c,d]{and Alberto Zaffaroni}
\affiliation[a]{Kavli IPMU (WPI), UTIAS, The University of Tokyo, Kashiwa, Chiba 277-8583, Japan}
\affiliation[b]{Institute for Nuclear Research and Nuclear Energy, Bulgarian Academy of Sciences, \\Tsarigradsko Chaussee 72, 1784 Sofia, Bulgaria}
\affiliation[c]{Dipartimento di Fisica, Universit\`a di Milano-Bicocca, I-20126 Milano, Italy}
\affiliation[d]{INFN, sezione di Milano - Bicocca, I-20126 Milano, Italy}
\emailAdd{morteza.hosseini@ipmu.jp}
\emailAdd{khristov@inrne.bas.bg}
\emailAdd{yuji.tachikawa@ipmu.jp}
\emailAdd{alberto.zaffaroni@mib.infn.it}

\abstract{We derive the general anomaly polynomial for a class of two-dimensional CFTs arising as twisted compactifications of a higher-dimensional theory on compact manifolds $\mathcal{M}_d$, including the contribution of the isometries of $\mathcal{M}_d$. We then use the result to perform a  counting of microstates for electrically charged and rotating supersymmetric black strings in AdS$_5\times S^5$ and AdS$_7\times S^4$ with horizon topology BTZ$ \ltimes S^2$ and BTZ$ \ltimes S^2 \times \Sigma_\mathfrak{g}$, respectively, where $\Sigma_\mathfrak{g}$ is a Riemann surface. We explicitly construct the latter class of solutions by uplifting a class of four-dimensional rotating black holes. We provide a microscopic explanation of the entropy of such black holes by using a charged version of the Cardy formula.}

\begin{document}

\setcounter{tocdepth}{2}
\maketitle

%*******************************************************************************
%
%     Main body
%
%*******************************************************************************

\date{Dated: \today}

% \hypersetup{
% colorlinks,breaklinks,
%             linkcolor=black
% }

% \tableofcontents

% \hypersetup{
% colorlinks,breaklinks,
%             linkcolor=[rgb]{0,0,0.7}
% }

\section{Introduction}
\label{sect:intro}

The charged Cardy formula, 
\be
  \log \rho (n , J) \approx  2 \pi \sqrt{\frac{c}{6} \left( n - \frac{c}{24} - \frac{J^2}{2k} \right)} \, ,
  \label{eq:intro:Charged-Cardy}
\ee
gives the asymptotic density of states $\rho (n , J)$ of a 2d conformal field theory (CFT) with a $\U(1)$ symmetry of level $k$
in a sector with a given $\U(1)$ charge. The above formula, and its natural generalization to multiple $\U(1)$ factors, has endless applications to the microscopic derivation of the Bekenstein-Hawking
entropy of asymptotically flat black holes or charged Ba\~nados-Teitelboim-Zanelli (BTZ) ones, where it has been used and derived many times
(see references \cite{Breckenridge:1996is,Maldacena:1997de,Dijkgraaf:2000fq,Kraus:2006nb,Ammon:2012wc,Montero:2016tif,Pal:2020wwd}
among many others).
In this paper we would like to apply it to the physics of asymptotically anti de Sitter (AdS) black strings in dimensions greater or equal to five.
We consider, in particular,   solutions interpolating between AdS$_{d+3}$ and AdS$_3\times \cM_d$, which can be seen either as black strings
or domain walls. They are holographically dual to  twisted compactifications of a $(d+2)$-dimensional CFT on $\cM_d$ that flow in the IR
to a two-dimensional CFT. Upon compactification on a circle with momentum along it, the black string becomes a black hole and the charged Cardy formula
gives a prediction for the corresponding entropy.%
\footnote{In what follows we use the convention of calling a {\it black string} any geometry with a near-horizon region containing a BTZ factor in its full ten- or eleven-dimensional form, in order to distinguish with the cases where the near-horizon only contains an AdS$_2$ factor. This is a non-standard terminology since it means that we denote as black strings all solutions that relate to a two-dimensional CFT, typically called black holes in the asymptotically flat literature.}

The CFTs obtained by twisted compactifications may have various $\U(1)$ symmetries, some of them arising from the flavor symmetries
of the original $(d+2)$-dimensional CFT, other coming  from the isometry group $G$  of the compactification manifold $\cM_d$.\footnote{The CFT may also have accidental symmetries. We will not consider this interesting case in this paper.} In the case of a sphere,
$\cM_2=S^2$, the latter is just the rotational symmetry of $S^2$. In order to use the charged Cardy formula, we need to determine the levels of the $\U(1)$ symmetries in the two-dimensional CFT. 
The latter can  be extracted  from  the anomaly polynomial of the 2d theory, which, in turn, can be obtained by integrating the anomaly polynomial of the higher-dimensional theory over $\cM_d$.  
All of these techniques are very standard and they can also be applied to non-Lagrangian theories, see e.g.~\cite{Benini:2009mz,Alday:2009qq,Bah:2011vv,Bah:2012dg,Bobev:2015kza,Kim:2017toz},
but the effect of the isometry of $\cM_d$ was not taken care of until recently.
When $\cM_d$ is a sphere, this generalization was performed in \cite{Bah:2019rgq,Bah:2019vmq}  using the Bott-Cattaneo  formula \cite{Bott-Cattaneo} and also extended to more complicated
compact manifolds. We will then make a further generalization by utilizing the connection to the  equivariant cohomology and the fixed point formula for the equivariant integration.
In principle, when abelian, the isometry $G$ can mix with the abelian flavor symmetries (and the R-symmetry in the supersymmetric case). We will present examples where this happens.

We then turn to the holographic side and compare the predictions of the charged Cardy formula with supergravity solutions of type IIB and M-theory.
We are interested, in particular, in adding charges and rotation to the supersymmetric black strings in AdS$_5\times S^5$, dual to ${\cal N}=4$ super Yang-Mills (SYM) compactified on $S^2$, and   in
AdS$_7\times S^4$, dual to the six-dimensional $\cN = (2,0)$ theory compactified on $S^2\times \Sigma_\fg$, found  in \cite{Benini:2013cda}. These static solutions  depend on a set of magnetic fluxes parameterizing the
inequivalent topological twists,  and the holographic central charge of the 2d $\cN = (0,2)$ CFT has been successfully
compared with the field theory predictions of $c$-extremization \cite{Benini:2012cz}. 

The general class of electrically charged and rotating black strings in AdS$_5\times S^5$ was constructed recently in \cite{Hosseini:2019lkt} and the entropy of the corresponding four-dimensional black hole matched
with the predictions of the refined topologically twisted index of ${\cal N}=4$ SYM \cite{Benini:2015noa}.%
\footnote{The refined topologically twisted index is supposed to compute the elliptic genus of the 2d CFT. See also \cite{Hosseini:2016cyf} for the evaluation of the topologically twisted index in the Cardy limit.}
The entropy, see \cite[(5.34)]{Hosseini:2019lkt} for example,  takes the form of the charged Cardy formula, although the connection was not noticed there.
In this paper we perform a further precision test,
explicitly comparing the levels of all symmetries including the rotational one.
We mostly work in the large $N$ limit, but we also observe that the  charged Cardy formula matches the results obtained from the high-temperature limit of the refined topologically twisted index of a 4d theory on $S^2$ also at finite $N$.

To proceed further, we consider asymptotically AdS$_7\times S^4$ supersymmetric black strings, corresponding to the compactification of the $(2,0)$ theory  on $S^2\times \Sigma_\fg$, and carrying  charge under the Cartan subgroup of the $\SO(5)$ R-symmetry in six dimensions and rotating along $S^2$. We explicitly 
construct  their near-horizon horizon geometry, which has  topology BTZ$ \ltimes S^2 \times \Sigma_\fg$, using a generalization of the methods introduced in  \cite{Hristov:2018spe}. We find the solution in the form of a black hole near-horizon geometry in the four-dimensional ${\cal N}=2$ supergravity that arises
as a dimensional reduction of seven-dimensional gauged supergravity on $\Sigma_\fg\times S^1$. The black hole carries $n$ units of momentum along the circle, and its entropy 
perfectly matches the charged Cardy formula using the levels for flavor and rotational symmetry predicted by the integration of the anomaly polynomial of the six-dimensional $(2,0)$ theory.

The very same near-horizon black hole geometry can be uplifted to five-dimensional gauged supergravity where it describes the horizon of an asymptotically AdS$_5$ black string. In this picture, the AdS$_5$
vacuum corresponds to the ${\cal N}=1$ superconformal theory obtained as the IR limit of the $(2,0)$ theory compactified on $\Sigma_\fg$   \cite{Bah:2011vv,Bah:2012dg} and the black string
describes its further compactification on $S^2$.

As usual with supersymmetric objects in AdS, supersymmetric black strings come with a non-trivial constraint among the possible electric charges allowed by supergravity. It is interesting to observe that, in all our examples, this constraint translates into the vanishing of the charge of the black string under the exact R-symmetry of the 2d CFT.  This is analogous to the expectations for magnetically charged black holes in AdS$_4$ \cite{Benini:2015eyy,Benini:2016rke}.

The paper is organized as follows. In section \ref{sec:Cardy} we briefly review the derivation of the charged Cardy formula.
In section \ref{sec:anomalies} we review the relevant facts about integrating anomaly polynomials across dimensions and we extend the analysis to the isometry group of the internal manifold, using the Bott-Cattaneo formula and equivariant integration.
We also provide various examples that are used in later sections or we find intriguing.
In section \ref{subsec:BSADS7} we construct a new class of dyonic rotating black strings  in AdS$_7\times S^4$ and we also review the analogous solutions in AdS$_5\times S^5$.
Finally, in section \ref{sec:microstate} we compare the supergravity results with the charged Cardy formula, using the levels computed in  section \ref{sec:anomalies}, finding perfect agreement.
We conclude in section \ref{sect:discussion} with comments and a discussion. In the  appendix we rephrase our supergravity findings in the language of attractors and entropy functions, using the gravitational blocks introduced in  \cite{Hosseini:2019iad}.

\section{Charged Cardy formula}\label{sec:Cardy} 

Here we review the derivation of the charged Cardy formula,
which tells the asymptotic density of states of a two-dimensional CFT with a $\U(1)$ symmetry
in a sector with a given $\U(1)$ charge.
This formula was found independently many times:
\eg\;it appears in the explanation of the entropy of asymptotically flat spinning black holes \cite{Breckenridge:1996is}; the influential paper by Maldacena, Strominger and Witten has it in \cite[Sec.~3.2]{Maldacena:1997de}, where the $\U(1)$ is clearly carried by chiral bosons
and its contribution can be explicitly taken care of;
the Farey tail paper also has it in \cite[(1.19)]{Dijkgraaf:2000fq}.
A more detailed discussion  in the context of holography in charged BTZ black holes
is in \cite{Kraus:2006nb}, which is then cited in a more recent review article \cite[Sec.~2.3]{Ammon:2012wc}.
The same formula was more recently rederived in \cite[Appendix B.2]{Montero:2016tif} in the context of the weak-gravity conjecture.
We will be brief; we mainly use our discussion here to set the notations.

We also note that the (uncharged) Cardy formula holds only after an order-1 averaging in the excitation number $n$, whose physical significance was recently emphasized in \cite{Mukhametzhanov:2019pzy}. 
The charged Cardy formula was also analyzed in this context in \cite{Pal:2020wwd}.
In this paper we will be naive and leave this averaging implicit.

\subsection{Non-supersymmetric case}\label{non-susy}

We start from the grand canonical partition function of a two-dimensional CFT coupled to a $\U (1)_{J}$ current on $T^2$, \ie\;
\be
 \label{CFT2:Z}
 Z = \tr_{\cH_{S^1}} \ex^{- \beta H + \ii \varepsilon P} \ex^{\ii \mu J} \, ,
\ee
where $H$ is the Hamiltonian, $P$ the  momentum along $S^1$ inside $T^2$, $J$ is the flavor symmetry charge,
$\beta$ is the inverse temperature;
$\varepsilon$ and $\mu$ are the  chemical potentials for $P$ and $J$, respectively.
The modulus of the torus is \be
 \label{tau:AdS}
2\pi \tau \equiv \varepsilon + \ii \beta\, .
\ee

Let us focus on the holomorphic part\footnote{%
The following analysis is very crude,
due to the very fact that there is no easy way to make the concept of the `holomorphic part of the partition function' very precise. 
This important caveat does not affect the derivation of the leading asymptotics.
} and write 
\bea
 Z 
 & = \int_{0}^{\infty} \! \rd n_l \int \rd J \? \rho(n_l , J) \ex^{2 \pi \ii \tau \left( n_l - \frac{c_l}{24} \right)   + \ii \mu J} \, ,
\eea
where $\rho(n_l,J)$ is the density of states.
We now need to recall the modular transformation property of the holomorphic part of the partition function.
\begin{equation}
Z\Big(\frac{a\tau+b}{c\tau + d}, \frac{\mu}{c\tau+d}\Big)
\sim \exp\left(
\frac{\pi\ii k}{(2\pi)^2}
\frac{c\mu^2}{c\tau+d} 
\right)
Z(\tau,\mu) \, ,
\label{Ztrans}
\end{equation}
where 
$k$ is the level of the $\U(1)$ current $J$,
and the symbol $\sim$ means we suppressed various important details coming from the fact that we repressed the anti-holomorphic dependence. 
The $k$-dependent exponential prefactor is known from long time ago, based on explicit examples. 
For a derivation which only uses general properties of 2d CFTs,  see \eg\;a discussion in \cite[Sec.~2]{Dyer:2017rul}.

In particular, we have \begin{equation}
 Z(\tau , \mu)=  
 \ex^{-\frac{ \ii \mu^2}{4\pi \tau} k }
 Z(- 1 / \tau , - \mu / \tau)
 \sim 
 \ex^{\frac{\ii \pi}{12 \tau} \left( c_l - \frac{3 \mu^2}{\pi^2} k \right)} \, ,
 \label{Zasymp}
\end{equation}
where $\sim$ is for the asymptotic form in the limit $\tau\to 0$
in which  $ Z(- 1 / \tau , - \mu / \tau)$ is dominated by the vacuum
and is given by $ \sim \ex^{\frac{\ii \pi c_l}{12 \tau}}$.
We neglected the part polynomial in $\tau$ which only gives a subleading correction.

We now use the inverse Laplace transform to read off $\rho(n_l,J)$:
\be
 \rho (n_l , J) = \int_\cC \rd \tau \? \rd \mu \? Z(\tau , \mu) \ex^{- 2 \pi \ii \tau \left( n_l - \frac{c_l}{24} \right)  - \ii \mu J} \, .
\ee
We replace $Z(\tau,\mu)$ by its asymptotic form \eqref{Zasymp}
\be
 \label{dmicro:n,cJ:S-tr}
 \rho(n_l , J) \approx \int_\cC \rd \tau \? \rd \mu \? \ex^{\frac{\ii \pi}{12 \tau} \left( c_l - \frac{3 \mu^2}{\pi^2} k \right)} \ex^{- 2 \pi \ii \tau \left( n_l - \frac{c_l}{24} \right) - \ii \mu J} \, .
\ee
The saddle point is at
\be
 \label{sp:tau0:omega0}
 \tau_{0} = \ii \sqrt{\frac{c_l}{24}\left( n_l - \frac{c_l}{24} - \frac{J^2}{2 k} \right)^{-1}} \, , \qquad
 \mu_0 =- \frac{2 \pi J}{k} \tau_0 \, .
\ee
Plugging back \eqref{sp:tau0:omega0} into \eqref{dmicro:n,cJ:S-tr} we find that
\be
\log  \rho (n_l , J) \approx 2 \pi \sqrt{\frac{c_l}{6} \left( n_l - \frac{c_l}{24} - \frac{J^2}{2k} \right)} \, .
\ee
Note that in the case of a single free boson $X$ whose current is $\partial X$, the operator of charge $J$ of the lowest dimension is the exponential operator $\ex^{iJX}$ and has the dimension $J^2/(2k)$.
Therefore the formula above simply says that the  excitation level in the charge $J$ sector should be thought of as $n_l-J^2/(2k)$.

Finally, putting back the anti-holomorphic part we obtain the \emph{charged Cardy formula} for a CFT$_2$ coupled to a $\U(1)_{J_l}$ current with level $k_l$
and a $\U(1)_{J_r}$ current with level $k_r$:
\be
\log  \rho (n_l , n_r , J_l , J_r) \approx 2 \pi \sqrt{\frac{c_l}{6} \left( n_l - \frac{c_l}{24} - \frac{J_l^2}{2 k_l} \right)} + 2 \pi \sqrt{\frac{c_r}{6} \left( n_r - \frac{c_r}{24} -\frac{J_r^2}{2k_r} \right)}  \, .
 \label{eq:Charged-Cardy}
\ee
The charged Cardy formula \eqref{eq:Charged-Cardy} can be trivially extended to the case of multiple left-moving and right-moving currents.

\subsection{Supersymmetric case and its relation to the anomaly polynomial}
\label{sec:susychargedcardy}

All examples in this paper are $(0,2)$ supersymmetric CFTs, with possibly multiple abelian symmetries $J_A$, whose background gauge fields we denote by $F^A$.
We will use conventions where supersymmetry is realized in the anti-holomorphic sector
and the 2d chirality matrix $\gamma_3$ is taken to be positive on anti-holomorphic fermionic movers. 
The 2d anomaly polynomial has the expansion
\be \cA_{2\rd} = \frac 12 {\cal A}_{AB} c_1(F^A) c_1(F^B) + \ldots \, ,
\label{2dk}
\ee
where the 't Hooft anomaly coefficients  are given by ${\cal A}_{AB}= \tr \gamma_3 J_A J_B$ in the case of Lagrangian theories.
We then define the level matrix $k_{AB}$ via
\be\label{signconventions} k_{AB} = - {\cal A}_{AB} \, .\ee
Notice the sign in \eqref{signconventions}. 
We choose it in such a way that the level matrix $k_{AB}$ in a unitary theory is positive definite for holomorphic currents. These are supported on
the non-supersymmetric side and  affect the density of states in the way discussed above.  Be aware that the symbol $k_{AB}$ is also widely used to denote the 't Hooft anomaly coefficients ${\cal A}_{AB}$ themselves, 
for example in \cite{Benini:2013cda}. We apologize for the possible sources of confusion. 

In this case we consider the elliptic genus:
\begin{equation}\label{eg}
 Z_\text{ell}(\tau,\mu^A) = \tr_{\cH_{S^1}} \ex^{2\pi \ii \tau \left(n_l-\frac{c_l}{24} \right)} (-1)^{F_r}  \ex^{\ii \mu^A J_A} \, .
\end{equation}
Here, all non-R currents are included in the currents $J_A$, irrespective of whether they are left-moving or right-moving,
and we regard all chemical potentials $\mu^A$ as holomorphic variables.

To derive the charged Cardy formula for the elliptic genus, 
we need its modular transformation law:
\begin{equation}\label{egt}
Z_\text{ell}\Big(\frac{a\tau+b}{c\tau + d}, \frac{\mu}{c\tau+d}\Big)
=(\epsilon_{a,b,c,d}){}^{c_r-c_l} \exp\left(
\frac{\pi\ii k_{AB}  }{(2\pi)^2}\frac{c\mu^A \mu^B}{c\tau+d} 
\right)
Z_\text{ell}(\tau,\mu^A) \, ,
\end{equation}
where $\epsilon_{a,b,c,d}$ is a multiplier system, which is a fixed universal non-trivial one-dimensional representation of $\SL(2,\bZ)$.
This very general transformation law can be checked for Lagrangian theories by localization \cite[(2.16)]{Benini:2013xpa}; we believe it is a universal property of elliptic genera.

The rest of the derivation is exactly the same as in the non-supersymmetric case.
We conclude that for supersymmetric states the charged Cardy formula is given by
\be
\log  \rho_\text{susy} (n_l,   J_A) \approx 2 \pi \sqrt{\frac{c_l}{6} \left( n_l - \frac{c_l}{24} - \frac 12 (k^{-1})^{AB} J_A J_B \right)} \, .
 \label{eq:susy:Charged-Cardy}
\ee
We stress  that $k_{AB}$ appearing here in the supersymmetric case can be read off from the anomaly polynomial.
Furthermore, $k_{AB}$ is not necessarily positive definite.

In the following we will consider 2d CFTs that arise as compactifications of a higher-dimensional theory on a manifold $\cM_d$. We will then use the symbols $J_i$ to denote symmetries arising from the isometries
of $\cM_d$ and $Q_i$  to denote  R-symmetries of the original   higher-dimensional theory. With a standard abuse of notation, we will also use the same letter to denote the charge of a state and the corresponding operator. 

\section{Anomaly polynomial and the dimensional reduction}\label{sec:anomalies}
The anomaly polynomial of a theory is a basic quantity which characterizes the change of the phase of its partition function when we perform gauge transformations of the background metric and/or the background gauge fields.
For a theory with an explicitly known Lagrangian, its anomaly polynomial can be easily computed using the standard formulae.
If no Lagrangian is known, the computation of the anomaly polynomial is more difficult, and requires various methods adapted to each situation. 

Luckily, if the theory under investigation is a dimensional reduction on a compactification manifold $\cM_d$ of a $D$-dimensional theory with a known anomaly polynomial $\cA_{D\rd}$, 
the anomaly polynomial $\cA_{(D-d)\rd}$ of the resulting $(D-d)$-dimensional theory can be easily computed by integrating over $\cM_d$.
In this paper, we are interested in particular to the cases when the compactification manifold $\cM_d$ has a continuous isometry group $G$.
Then the lower-dimensional theory has $G$ as an additional flavor symmetry,
and the lower-dimensional anomaly polynomial $\cA_{(D-d)\rd}$ should include the curvature of the rotational symmetry $G$.

When $\cM_d$ is an even dimensional sphere, this computation can be performed using the Bott-Cattaneo formula \cite{Bott-Cattaneo}, as was demonstrated in \cite{Bah:2019rgq,Bah:2019vmq}.
The aim of this section is to review this computation, and to make a further generalization to the case when $\cM_d$ is not a sphere, by using the equivariant cohomology and the fixed point formula for the equivariant integration.
In this section, we use the convention that $a\rd:=a+2$ such that the subscript of the anomaly polynomial is its degree as a differential form.

\subsection{Generalities}
Let us start by recalling the case when we do not care about the isometry of $\cM_d$.
The relation between the higher-dimensional anomaly polynomial $\cA_{D\rd}$
and the lower-dimensional one $\cA_{(D-d)\rd}$ is a well-known one, \begin{equation}
\cA_{(D-d)\rd} = \int_{\cM_d} \cA_{D\rd} \, .\label{eq:basic-anom-reduction}
\end{equation}
For Lagrangian theories this relation was known from time immemorial;
it was first applied to non-Lagrangian theories in \cite[Sec.~3.5]{Benini:2009mz} and \cite{Alday:2009qq}.

To understand the generalization to include the background gauge field for the isometry of $\cM_d$, 
it is instructive to recall how the relation \eqref{eq:basic-anom-reduction} is derived. 
For this purpose we need to recall how the anomaly polynomial $\cA_{D+2}$ encodes the anomaly in the first place.
Let $X_{D}$ be the spacetime on which the theory lives.
We then let $Y_{D+1}=X_{D}\times S^1$.
To obtain the change in the phase of the partition function under a combination of a diffeomorphism and a gauge transformation,
we introduce the background metric and the background gauge fields on $Y_{D+1}$ such that $Y_{D+1}$ is obtained by 
starting from $X_{D}\times [0,1]$ and gluing two boundaries $X_{D}|_0$ and $X_{D}|_1$ by the said diffeomorphism and the gauge transformation.
Then the change in the phase of the partition function is \begin{equation}
\int_{Y_{D+1}} \CS_{D+1},\qquad \text{where}\qquad \rd\,\CS_{D+1} = \cA_{D\rd}\, .
\end{equation} 

That we compactify on $\cM_d$ means that we take $X_D=X_{D-d} \times \cM_d$,
and therefore we have $Y_{D+1}=Y_{D-d+1} \times \cM_d$.
Then we have \begin{equation}
\int_{Y_{D+1}} \CS_{D+1} = \int_{Y_{D-d+1}} \CS_{D-d+1},\qquad
\text{where}\qquad  \CS_{D-d+1}=\int_{\cM_d} \CS_{D+1}\, .
\label{integration}
\end{equation} 
Let us now take a manifold $Z_{D-d+2}$ such that $\partial Z_{D-d+2}=Y_{D-d+1}$,
and set $Z_{D+2}=Z_{D-d+2}\times \cM_d$.
The relation \eqref{integration} then implies that 
$\cA_{(D-d)\rd}$ and $\cA_{D\rd}$  always satisfy \begin{equation}
\int_{Z_{D+2}} \cA_{D+2} = \int_{Z_{D-d+2}} \cA_{D-d+2}, \qquad
\text{which means}\qquad  \cA_{D-d+2}=\int_{\cM_d} \cA_{D+2} \, ,
\end{equation}
which is the basic relation \eqref{eq:basic-anom-reduction}.

This analysis makes it clear that to include the effect of the background gauge field for the isometry group $G$ of $\cM_d$,
we need to take $Z_{D+2}$ to be a nontrivial $\cM_d$ bundle over $Z_{D-d+2}$ with a nontrivial $G$ connection, so that $Z_D$ has the form
\begin{equation}
 \cM_d\to Z_{D+2} \to Z_{D-d+2}\, .
\end{equation}
For notational simplicity, we will use $E=Z_{D+2}$ and $B=Z_{D-d+2}$ below, so that the fibration is
\begin{equation}
 \cM_d \to E\to B\, , \label{eq:fibered}
\end{equation} where $\dim E=\dim B+d$.

For example, consider the case when $\cM_d=S^2$, whose $\SO(3)$ isometry becomes a flavor symmetry in the lower-dimensional theory. 
Suppose furthermore that the higher-dimensional theory has a $\U(1)$ symmetry,
and that  we include $n$ fluxes of it through $S^2$.
Then, on $E$, we have a background $\U(1)$ gauge field $F$ which satisfy \begin{equation}
\int_{S^2} \frac{F}{2\pi} = n\, .\label{eq:fibered-u1}
\end{equation}
When $E$ is a direct product $B\times S^2$, such a gauge field $F$ has non-zero components only along the $S^2$ direction.
But when the $\SO(3)$ gauge field is nontrivial, $F$ necessarily has non-zero components also along the $B$ direction.
Therefore, integrals of the form \begin{equation}
\int_{S^2} \Big(\frac{F}{2\pi} \Big)^n \, ,
\end{equation} can result in nontrivial differential forms on $Z_D$,
describing the anomaly of the $\U(1)$ symmetry already existent in the higher-dimensional theory 
and of the $\SO(3)$ symmetry arising from the isometry of the compactification manifold $S^2$.

As another example,  let us analyze how to study the effect of the gravitational part of the higher-dimensional anomaly.
In this case, one needs to study the Pontryagin classes $p_i(TE)$.
In a fiber bundle \eqref{eq:fibered}, we can split the tangent bundle of the total space into the subbundle along $B$ and $\cM_d$, respectively. 
Let us write this decomposition as \begin{equation}
TE = \pi^*(TB)\oplus \cT_{E/B} \, , \label{eq:ds}
\end{equation} where the first summand is obtained by pulling back the tangent bundle of $B$ via the fiber projection $\pi:E\to B$,
and the second summand $\cT_{E/B}$ is defined to be its orthogonal complement,
which is called the relative tangent bundle.
The  decomposition \eqref{eq:ds} above allows us to rewrite $p_i(TE)$ in terms of $p_i(TB)$ and $p_i(\cT_{E/B})$.

For example, when $\cM_d=S^2$, $\cT$ can be thought of as an $\SO(2)\simeq \U(1)$ bundle over $E$ satisfying \eqref{eq:fibered-u1} with $n=2$, the Euler number of $S^2$.
Note that $\cT$ equals $T\cM_d$ when restricted on a single fiber $\cM_d$;
we abuse the notation slightly and simply use $T\cM_d$ for $\cT_{E/B}$.

Summarizing, $\cA_{D-d+2}$ can be evaluated 
if we know how to describe the cohomology groups of the total space $E$ in terms of the cohomology groups of $B$ and $\cM_d$, 
and if we know how to integrate cohomology classes on $E$ over $\cM_d$.
We describe two methods to achieve this goal, depending on the type of manifolds $\cM_d$.
They are i) the formula of Bott-Cattaneo when $\cM_d=S^{2k}$,
and ii) the equivariant integration of  equivariant cohomology groups in the general case.

\subsubsection[Bott-Cattaneo formula for \texorpdfstring{$S^{2k}$}{S**(2k)}]{Bott-Cattaneo formula for $S^{2k}$}

Let us say $V$ is a real $(2k+1)$-dimensional vector bundle over $B$.
Taking the unit sphere at each fiber, 
we have an $S^{2k}$ bundle $E$ over $B$, where $d=2k$.\footnote{%
Here it is important that we have a bundle of even-dimensional spheres.
The odd-dimensional spheres behave rather differently, since $e$ contains a Chern-Simons term, rather than a characteristic class, of the isometry connection.
See e.g.~\cite{Kim:2012wc,Bah:2020jas} for discussions.
}
Split the tangent bundle as in \eqref{eq:ds},
and let $e$ be the Euler class of $E$. 
We have $\int_{S^{2k}}e=2$, the Euler number of $S^{2k}$.
The formula of Bott-Cattaneo \cite[Lemma~2.1]{Bott-Cattaneo} is as follows: \begin{equation}
\int_{S^{2k}} e^{2s+1} = 2(p_k(V))^s\, , \qquad
\int_{S^{2k}} e^{2s} = 0\, .
\label{eq:Bott-Cattaneo}
\end{equation}

To the authors' knowledge, this formula was first used in hep-th in \cite{Freed:1998tg,Harvey:1998bx}, in the context of the anomaly cancellation of the R-symmetry part of the M5-brane.
The first application to the isometry of the compactification manifold was performed in \cite{Bah:2019rgq,Bah:2019vmq}, see in particular \cite[Sec.~5.1]{Bah:2019rgq}.

We also note that the original proof in \cite[Lemma~2.1]{Bott-Cattaneo} was a simple application of the splitting principle, and was done at the level of cohomology;
those who prefer   the discussion at the level of the differential form can find it in appendices of \cite{Freed:1998tg,Harvey:1998bx}.

\subsubsection{Equivariant integration}
\label{sect:equiv:int:anomaly}

Let us discuss next a method which is applicable to a larger class of compactification manifolds $\cM$ with isometry group $G=\U(1)^n$.
We note that our computation involves only a combination of Pontryagin classes of the spacetime manifold,  the characteristic polynomials of background gauge fields in higher dimensions, and the background gauge fields for the isometry of $\cM$.
The cohomology classes involved are universal and independent of the specific choice of $E$ and $B$, in a sense which can be made mathematically precise. 
Such universal cohomology classes make up what is known as  the $G$-equivariant cohomology group%
\footnote{There are many different constructions of equivariant cohomology groups which eventually give isomorphic groups, just as there are many different constructions of ordinary cohomology groups.
The description we are using here is the one $H_G^\bullet(\cM):=H^\bullet((\cM \times EG)/G)$, 
where $G\to EG\to G$ is the universal $G$-bundle over the classifying space $BG$ of $G$. 
Then $(\cM\times EG)/G$ is the associated universal $\cM$ bundle over $BG$,
the cohomology over $\bR$ of which can be constructed from the cohomology of $\cM$ and the cohomology of $BG$.
Finally, the cohomology of $BG$ can be identified with the characteristic classes of $G$-bundles via the Chern-Weil homomorphism.

One can use a more concrete description of equivariant cohomology using the differential forms, the exterior differential $d$, and the interior product $\iota_k$ by the isometry $k$.
This approach was reviewed in great detail in \cite{Cordes:1994fc}, 
and also more recently in \cite[Appendix B]{Bah:2019rgq} in the context of the reduction of the anomaly polynomial on a compactification manifold.
}
which is denoted by $H_G^\bullet(\cM)$.

For many manifolds, including the toric K\"ahler manifolds we will consider below,
we simply have \begin{equation}
H_{\U(1)^n}^\bullet(\cM)= H^\bullet(\cM) \otimes \bR[c_1(J_1),\ldots, c_1(J_n)] \, ,
\end{equation}
where $c_1(J_i)$ is the 1st Chern class along the base $B$ of the background gauge field of the $i$-th $\U(1)$ isometry.
Let us further assume that the only discrete isolated points on $\cM_d$ are fixed under the $\U(1)^n$ isometry action.
Then the integral of equivariant cohomology classes can be done most conveniently by the localization formula
\begin{equation}
 \int_{\cM} \omega = \sum_{\text{$p$:fixed points}} \frac{\omega|_p}{e(T\cM|_p)} \, ,
\end{equation}
where the sum is over all fixed points $p$,
$\omega|_p$ and $T\cM|_p$ are the restriction of $\omega$ and $T\cM$ at $p$,
and $e$ is the Euler class.
To compute the Euler class, we regard $T\cM|_p$ as a complex vector bundle,
which we write as a sum of complex line bundles: $T\cM|_p = \bigoplus_j L_j$.
Let us say $L_j$ transforms under a definite charge $(q_1^{(j)},\ldots,q_n^{(j)})$ under $\U(1)^n$. Then 
\begin{equation}
e(T\cM|_p)=\prod_j c_1(L^{(j)})=\prod_j \sum_i q_i^{(j)} c_1(J_j)\, .
\end{equation}

Let us use this machinery to reproduce the Bott-Cattaneo formula.
Take $\cM=S^{2k}$ with its standard $\U(1)^k$ action.
There are two fixed points, the north pole and the south pole.
The Euler class $e(TS^{2k})$ restricts to $\pm \prod c_1(J_i)$, where $c_1(J_i)$ is the first Chern class for the $i$-th $\U(1)$ isometry.
The localization formula then leads to \begin{equation}
\int_{S^{2k}} \ex^{n} = \frac{(+\prod_i c_1(J_i))^n}{+\prod_i c_1(J_i)} +\frac{(-\prod_i c_1(J_i))^n}{-\prod_i c_1(J_i)}
= \begin{cases}
2 (\prod_i c_1(J_i))^{2s} & (n=2s+1)\, , \\
0 & (n=2s)\, .
\end{cases}
\end{equation}
This reproduces \eqref{eq:Bott-Cattaneo},
since $p_k(V)=\prod_i c_1(J_i)^2$.

\subsection{Going from 4d to 2d}
Let us now apply the general machinery explained above in a few specific cases.
The first case we analyze is the compactification of 4d theories on $S^2$ to 2d theories.
Suppose there is a $\U(1)$ symmetry (which we call an R-symmetry) in 4d, with the anomaly polynomial \begin{equation}
\cA_{4\rd} = \frac{\tr R^3}6 c_1(R)^3 -\frac{\tr R}{24}c_1(R) p_1(TZ_6)\, .
\end{equation}
Here, as usual, $\tr$ is the trace over the label of Weyl fermions in the case of a Lagrangian theory.
We compactify it on a round $S^2$, with the $\U(1)_R$ flux $n$.
We would like to determine the anomaly of the 2d theory, including the $\SO(3)$ symmetry rotating $S^2$. 
Let us call the $\U(1)$ subgroup of $\SO(3)$ as $\U(1)_J$, with the normalization that the spin-1 representation has eigenvalues $-1,0,1$.

The manifolds involved form the fibration $S^2\to Z_6\to Z_4$,
and $TZ_6=TZ_4 \oplus TS^2$. Therefore 
\begin{equation}
p_1(TZ_6)=p_1(TZ_4)+e(TS^2)^2.
\end{equation} As we put $n$ units of flux on $S^2$, we have \begin{equation}
c_1^{4\rd}(R)=c_1^{2\rd}(R)+n\, e(TS^2)/2\, .
\end{equation}
Note that $e(TS^2)$ is the Euler class so it integrates to $2$ on $S^2$.

Integrating over $S^2$ using the Bott-Cattaneo formula \eqref{eq:Bott-Cattaneo},
we find \begin{equation}
\cA_{2\rd}=\int_{S^2}\cA_{4\rd}=\frac{n \tr R^3}{2} c_1(R)^2
+ \frac{ n^3 \tr R^3 - n\tr R}{24} c_1(J)^2 
-\frac{n\tr R}{24} p_1(TZ_4)\, .
\label{eq:4d2d}
\end{equation}

We note that the $\U(1)$ current algebra of level $k$ corresponds to a term in the anomaly polynomial $\pm\frac{k}2 c_1(J)^2$, where the sign depends on whether the current is right-moving or left-moving, respectively, see \eqref{signconventions}.
Similarly, the $\SU(2)$ current algebra of level $k_{\SU(2)}$ corresponds to a term in the anomaly polynomial of the form $\pm \frac{k_{\SU(2)}}2 \tr (F/(2\pi))^2$.
We are taking a $\U(1)_J\subset \SU(2)$ subgroup where the doublet has $\U(1)$ charge $\pm1/2$, and therefore this reduces to $\pm\frac{k_{\SU(2)}}4 c_1(J)^2$.

\subsubsection{Free chiral fermions}

Let us check this result when the 4d theory is a chiral fermion of $\U(1)$ charge $1$.
The zero modes on $S^2$ with $n$ units of flux form an irreducible representation of $\SU(2)$ of dimension $|n|$ 
and their chirality is determined by the sign of $n$.
For simplicity let us assume $n\ge 0$.
Then the 2d theory consists of $n$ complex chiral fermion whose $J$ charges are $(n-1)/2$, $(n-3)/2$, \ldots, $(1-n)/2$.
As a charge $q$ fermion contributes $(q/2)c_1(J)^2$ to the anomaly, the term proportional to $c_1(J)^2$ in the anomaly polynomial should be \begin{equation}
\frac12\left[\left(\frac{n-1}2\right)^2 + \left(\frac{n-1}2-1\right)^2 + \cdots + \left(\frac{1-n}2\right)^2 \right] c_1(J)^2
= \frac{n^3-n}{24} c_1(J)^2\, ,
\end{equation} agreeing with \eqref{eq:4d2d}.

\subsubsection{General \Nequals1 theories and the charged Cardy formula}

When the 4d theory is \Nequals1 supersymmetric with a non-anomalous integer $\U(1)_R$ symmetry, 
we can preserve 2d \Nequals{(0,2)} supersymmetry by turning on a single unit $n=-1$ of $\U(1)_R$ flux through $S^2$.
The anomaly in 2d can be obtained simply by setting $n=-1$ in \eqref{eq:4d2d}.
For example, from the coefficient of the $c_1(J)^2$ term, using \eqref{2dk} and \eqref{signconventions}, we find
\begin{equation}
 k = \frac{1}{12} ( \tr R^3-\tr R ) \, .
\end{equation}

In the case where the four-dimensional $\cN=1$ theory is superconformal and $R$ is the exact R-symmetry,%
\footnote{Notice that this case, called the universal twist \cite{Benini:2012cz,Benini:2015bwz}, is not particularly interesting for our purposes.
It leads to a $2\rd$ unitary CFT $(c_r > 0)$, at large $N$, only when the compactification is done on higher genus Riemann surfaces.
As it can be seen from  \eqref{eq:universal}, in the case of $S^2$ at large $N$, where $a_{4\rd} = c_{4\rd}$, the central charge $c_r$ is negative.
Moreover, in many examples the four-dimensional R-symmetry is not integer-valued.}
we obtain
\begin{equation}
 k = \frac{8}{27}(3c_{4\rd}-2a_{4\rd}) \, ,
 \label{eq:SU(2)level}
\end{equation}
where $c_{4\rd}$ and $a_{4\rd}$ are the central charges \cite{Anselmi:1997am}.
Also, if the 2d R-symmetry can be directly identified with the superconformal R-symmetry, the 2d anomaly polynomial has the form
\begin{equation}
 \cA_{2\rd}= \frac{c_r}{6} c_1(R)^2 + \frac{c_l-c_r}{24} p_1(TZ_4)\, .
 \label{eq:2danom}
\end{equation}
Comparing with \eqref{eq:4d2d} with $n=-1$, one finds 
\bea
\label{eq:universal}
c_r&= -3 \tr R^3=-\frac{16}{3}(5a_{4\rd} -3 c_{4\rd})\, ,\\
c_l&= - ( 3\tr R^3 -\tr R )=-\frac{32}{3}a_{4\rd}\, .
\eea

In a general gauge theory with abelian flavor symmetries we have many choices of R-symmetries and each corresponds to a different twisted compactification.
The quantities in \eqref{eq:SU(2)level} are then replaced by
\bea
 \label{a:c:trial}
  a_{4\rd} ( \fs_I ) = \frac{9}{32} \tr R^3 - \frac{3}{32} \tr R \, , \qquad c_{4\rd}( \fs_I ) = \frac{9}{32} \tr R^3 - \frac{5}{32} \tr R \, ,
\eea
where
\bea
 \tr R^3  &=   \text{ dim}\, G + \sum_{I} \text{ dim}\,\fR_I  ( \fs_I  - 1 )^3 \, , \\
 \tr R &=   \text{ dim}\, G + \sum_{I} \text{ dim}\,\fR_I  ( \fs_I  - 1 ) \, .
\eea
Here, we are considering a gauge theory with gauge group $G$ and chiral matter fields in representation $\fR_I$ with integer R-charge $\fs_I$.
Moreover, the exact 2d R-symmetry is different from the 4d one and thus one needs to perform the $c$-extremization \cite{Benini:2012cz,Benini:2013cda} to find it.
We will give an explicit example in section \ref{sect:N=4SYM}.

In all cases we can apply the charged Cardy formula \eqref{eq:susy:Charged-Cardy} and we find
\be
 \label{N=(0,2):dmicro}
 \log \rho (n_l , n_r , J) \approx 2 \pi \sqrt{\frac{c_l}{6} \left( n_l - \frac{c_l}{24} - \frac{27 J^2}{16 \left( 3 c_{4\rd} ( \fs_I ) - 2 a_{4\rd} (\fs_I) \right)} \right)} \, ,
\ee
which remarkably agrees with \cite[(5.7)]{Hosseini:2019lkt}, where the density of states has been extracted in the Cardy limit and at finite $N$ from the refined topologically twisted index \cite{Benini:2015noa} which is
supposed to compute the equivariant elliptic genus of the 2d CFT.%
\footnote{With the identification  $e_0=n_l - \frac{c_l}{24}$.}  Also, the position of the saddle point \eqref{sp:tau0:omega0},
when we substitute the level \eqref{eq:SU(2)level},
agrees with \cite[(5.6)]{Hosseini:2019lkt}.

\subsubsection[\texorpdfstring{$\cN= 4$}{N} super Yang-Mills in the large \texorpdfstring{$N$}{N} limit]{\Nequals4 super Yang-Mills in the large $N$ limit}
\label{sect:N=4SYM}

As a concrete example, let us put \Nequals4 super Yang-Mills theory on $S^2$.
We use a basis of the $\U(1)^3\subset \SO(6)_R$ symmetry assigning charge $+1$ to chiral superfields $\Phi_{1,2,3}$, respectively;
we call their generators and field strengths as $Q_{1,2,3}$ and $F_{1,2,3}$, respectively.
The 4d anomaly polynomial in the large $N$ limit, for the gauge group $\U(N)$, is
\begin{equation}
 \cA_{4\rd} \approx \frac{N^2}2 c_1(F_1)c_1(F_2)c_1(F_3)\, .\label{eq:4dN=4SYM}
\end{equation}

Let us now embed the 2d $\U(1)_R$ symmetry in the direction $\Delta_i Q_i$ with $\sum \Delta_i=2$.
We write $c_1(F_i)=\Delta_i c_1^{2\rd}(F_R) - (\fs_i/2) (e/2)$ where $e$ is the Euler class of $S^2$,
so that  the flux on $S^2$ is given by $-\fs_i/2=\int_{S^2} c_1(F_i)$.
Supersymmetry in 2d requires $\sum\fs_i=2$.
Plugging this into \eqref{eq:4dN=4SYM} and integrating over $S^2$ using the Bott-Cattaneo formula, one finds
\be
 \cA_{2\rd} \approx - \frac{N^2 }{2} (\Delta_1 \Delta_2 \fs_3 + \Delta_2 \Delta_3 \fs_1 + \Delta_3 \Delta_1 \fs_2 ) c_1(F_R)^2  - \frac{N^2}{8} \fs_1 \fs_2 \fs_3 \? c_1 ( J ) ^2 \, .
 \label{eq:4dN=4SYMto2d}
\ee
We can now  extract the 2d trial central charge using \eqref{eq:2danom};
this parameterizes the mixing of the R-symmetry with the  flavor symmetries.
We can also read off the level of the rotational symmetry along $S^2$
using \eqref{2dk} and \eqref{signconventions}.
The results are:
\bea
 \label{scft:ckN4}
c_r(\Delta_i) &= - 3N^2  (\Delta_1 \Delta_2 \fs_3 + \Delta_2 \Delta_3 \fs_1 + \Delta_3 \Delta_1 \fs_2 ) \, , \\
k &= \frac{N^2}{4} \fs_1 \fs_2 \fs_3 \, .
\eea
Note that,
\be
 k = \frac{8}{27}\left( 3c_{4\rd} (\fs_i) -2a_{4\rd} (\fs_i) \right) \, ,
\ee
where
\be
  a_{4\rd} ( \fs_i) = c_{4\rd} ( \fs_i) = \frac{9}{32} \tr R^3 = \frac{9N^2}{32} \Big( 1 + \sum_{i = 1}^3 (\fs_i - 1)^3 \Big) \, ,
\ee
since $\tr R$ is identically zero  for $\cN=4$ SYM.
The exact central charge of the 2d theory can be obtained by extremizing $c_r ( \Delta_i )$ with respect to $\Delta_i$ with the constraint $\sum_{i=1}^3 \Delta_i=2$, and reads
\cite{Benini:2013cda}
\be
c_{\text{CFT}} =%= \frac{ 3 R_{{\rm AdS}_3}}{2 G_N^{(3)}} = %\frac{ 3 \ex^{ f(r_0) + 2 g(r_0)} {\rm vol} (\Sigma_\fg) }{2 G_N^{(5)}}  = 
  12 N^2 \frac{ \fs_1 \fs_2 \fs_3}{ \fs_1^2+\fs_2^2+\fs_3^2 - 2 \fs_1\fs_2 - 2 \fs_2 \fs_3 -2 \fs_3\fs_1} \, .
\ee

\subsection{Going from 6d to 2d}
Let us next discuss the compactification of 6d \Nequals{(2,0)} theory on four-dimensional manifolds. 
The eight-form anomaly polynomial of the abelian six-dimensional $\cN=(2,0)$ theory is given by \cite{Witten:1996hc}
\be
 \cA_{6\rd} [1] = \frac{1}{48} \left[ p_2 (R) - p_2 (T Z_8) + \frac{1}{4} \left( p_1 (T Z_8) - p_1 ( R ) \right)^2 \right] .
\ee
Then, for a generic $(2,0)$ theory of type $G$ it reads
\be
 \label{A6d:full}
 \cA_{6\rd} [G] = r_G \cA_{6\rd} [1] + \frac{d_G h_G}{24} p_2 ( R ) \, ,
\ee
where $r_G$, $d_G$, and $h_G$ are, respectively, the rank, dimension, and Coxeter number of $G$.
We restrict out attention to $G = A_{N-1}$ and the leading piece in the large $N$ limit,
which has the form
\be
 \label{A6d:largeN}
 \cA_{6\rd} [A_{N-1}] \approx \frac{N^3}{24} p_2 ( R ) \, ,
\ee
as is obvious from \eqref{A6d:full}.

\subsubsection[\texorpdfstring{$(2,0)$}{(2,0)} theory on \texorpdfstring{$S^2\times \Sigma_\fg}{S**2 x Sigma(g)}$ in the large \texorpdfstring{$N$}{N} limit]{$(2,0)$ theory on $S^2\times \Sigma_\fg$ in the large $N$ limit}
\label{sec:S2Sigma}

Let us compactify the six-dimensional theory on $S^2 \times \Sigma_\fg$,
where $\Sigma_\fg$ is a Riemann surface of genus $\fg$. We will consider the gravity dual of this case in section \ref{subsec:BSADS7}.
The holonomy group is $\SO(2) \times \SO(2)$ and we can preserve $\cN=(0,2)$ supersymmetry in two dimensions
by turning on an abelian background gauge field coupled to an $\SO(2) \times \SO(2) \subset \SO(5)_R$, embedded block-diagonally.
We only consider the case that  the $\SO(5)$ R-symmetry bundle is a sum of two line bundles whose first Chern classes are $x_{1,2}$, so that
\be
 p_1 ( R ) = x_1^2 + x_2^2 \, , \qquad p_2 ( R ) = x_1^2 x_2^2 \, .
\ee
We parameterize the fluxes through $S^2$ and $\Sigma_\fg$ by
\begin{equation}
 - \ft_\sigma = \int_{S^2} x_\sigma\, ,\qquad
 - \fs_\sigma = \int_{\Sigma_\fg} x_\sigma\, .
\end{equation}
In addition, we embed the 2d trial R-symmetry to the 6d R-symmetry using parameters $\Delta_\sigma$.
In the end we perform the replacement
\be
 x_\sigma \to - \frac{\ft_\sigma}{2} e(S^2) - \frac{\fs_\sigma}{2(1-\fg)} e(\Sigma_\fg) + \Delta_\sigma c_1 ( F_R ) \, , \qquad \sigma = 1, 2 \, ,
\ee
where $e(S^2)$ and $e(\Sigma_\fg)$ are the Euler classes of the respective surfaces.
The parameters satisfy the following constraints to preserve the supersymmetry:
\be
 \Delta_1 + \Delta_2 = 2 , \qquad \ft_1 + \ft_2 = 2 \, , \qquad \fs_1 + \fs_2 = 2(1-\fg) \, .
\ee

A short computation using the Bott-Cattaneo formula gives
\bea
 \label{eq:A2d:fromA6d:largeN2}
  \cA_{2\rd} & \approx \frac{N^3}{12} \left (  \fs_1 \ft_1 \Delta_2^2 + 2( \fs_1\ft_2+\fs_2 \ft_1) \Delta_1 \Delta_2 +  \fs_2 \ft_2 \Delta_1^2\right )  c_1( F_R )^2 \\
 & \qquad + \frac{N^3}{48}    \ft_1 \ft_2 (\fs_1\ft_2+\fs_2 \ft_1)  c_1( J )^2  \, ,
\eea 
where  $c_1(J)$ is the first Chern class of the background $\U(1)$ gauge field coupled to the rotation of $S^2$.
We note that the first line was already computed in \cite[(C.10), (C.11)]{Hosseini:2018uzp},
while the second line is the correction to the anomaly polynomial due to the angular momentum along $S^2$.
We immediately extract the 2d trial central charge and, using \eqref{2dk} and \eqref{signconventions}, the level of the rotational symmetry along $S^2$
\bea
\label{scft:ck(2,0)}
c_r(\Delta_i) &= \frac{N^3}{2} \left (  \fs_1 \ft_1 \Delta_2^2 + 2( \fs_1\ft_2+\fs_2 \ft_1) \Delta_1 \Delta_2 +  \fs_2 \ft_2 \Delta_1^2\right ) \, , \\
k &= - \frac{N^3}{24}  \ft_1 \ft_2 (\fs_1\ft_2+\fs_2 \ft_1)\, .
\eea
The exact central charge of the 2d theory can be obtained by extremizing $c_r ( \Delta_i )$ with respect to $\Delta_i$  with the constraint $\sum_{i=1}^2 \Delta_i=2$ and reads
\cite{Benini:2013cda}
\be
c_{\text{CFT}} =%= \frac{ 3 R_{{\rm AdS}_3}}{2 G_N^{(3)}} = %\frac{ 3 \ex^{ f(r_0) + 2 g(r_0)} {\rm vol} (\Sigma_\fg) }{2 G_N^{(5)}}  = 
  2 N^3 \frac{ \fs_1^2 \ft_2^2 + \fs_1 \fs_2 \ft_1 \ft_2 +\fs_2^2 \ft_1^2}{ \fs_1 (2 \ft_2 -\ft
 _1) + \fs_2 (2 \ft_1 -\ft
 _2)} \, .
\ee

\subsubsection[\texorpdfstring{$(2,0)$}{(2,0)} theory on toric \texorpdfstring{K\"ahler}{Kahler} surfaces]{$(2,0)$ theory on toric K\"ahler surfaces}

Let us next consider the 6d theory on compact toric K\"ahler surfaces $\cM$.
We first summarize the mathematical information we need.
Such a complex surface has $\U(1)^2$ isometry under which the fixed points are isolated,
and is specified by the toric data $\vec n_\ell \in \bZ^2$.
Here, $\ell=1,2,\ldots,n$ where $n$ is the number of the fixed points,
and we assume that the vectors $\vec n_\ell$ are ordered counter-clockwise;
as a convenience, we regard the subscripts are defined modulo $n$.
We only consider the case when $\cM$ is smooth, for which we have $\pair{\vec n_{\ell}}{\vec n_{\ell+1}}=1$.
% where we use the notation
%\begin{equation}
% \pair{\vec a}{\vec b}=\det(\vec a,\vec b).
%\end{equation}
Each vector $\vec n_\ell$ specifies a divisor $D_\ell$ at which a linear combination of two $\U(1)$ isometries specified by $\vec n_\ell$ degenerates.
Then $D_\ell$ and $D_{\ell+1}$ intersect and specify the $\ell$-th fixed point $x_\ell$.

We denote two first Chern classes of the $\U(1)^2$ isometry by $\epsilon_{a}=c_1(J_a)$ for $a=1,2$.
The tangent bundle at the fixed point $x_\ell$ splits as a sum of two line bundles whose first Chern classes are $\epsilon^{(\ell)}_{1,2}$, which are given by
\begin{equation}
\epsilon_1^{(\ell)} = -\pair{\vec n_{\ell+1} }{\vec\epsilon}\, ,\qquad
\epsilon_2^{(\ell)} = \pair{\vec n_{\ell} }{\vec\epsilon} \, ,
\end{equation}
in terms of $\vec\epsilon=(\epsilon_1,\epsilon_2).$
The equivariant localization formula now reads \begin{equation}
\int_{\cM} \omega = \sum_\ell\frac{\omega|_{x_\ell}}{\epsilon_1^{(\ell)}\epsilon_2^{(\ell)}}
= \sum_\ell\frac{\omega|_{x_\ell}} {  \pair{\vec n_{\ell} }{\vec\epsilon}\pair{\vec\epsilon }{\vec n_{\ell+1} } }\, .
\label{toric-equiv-integral}
\end{equation}

We note that $H^2(\cM)$ is $(n-2)$-dimensional, while $H^2_{\U(1)^2}(\cM)$ is extended by the first Chern classes of the $\U(1)^2$ isometry and therefore is $n$-dimensional.
The natural basis elements are  given by $c_1(L_\ell)$,  the 1st Chern classes of the equivariant line bundles $L_\ell$ corresponding to the divisor $D_\ell$.
Among them, purely equivariant bundles specified by $\vec w\in \bZ^2$ correspond to \begin{equation}
\sum_\ell \pair{\vec w}{\vec n_\ell} c_1(L_\ell)\, .\label{pure}
\end{equation}

The restriction of $c_1(L_\ell)$ to the fixed point $x_{\ell'}$ is given by
\begin{equation}
c_1(L_\ell)|_{x_{\ell'}} = \left\{
\begin{array}{lcr@{\quad}l}
\epsilon_2^{(\ell-1)} &=& \pair{\vec n_{\ell-1} }{\vec\epsilon} & \text{if $\ell'=\ell-1$}\, , \\
\epsilon_1^{(\ell)}  &= & -\pair{\vec n_{\ell+1} }{\vec\epsilon} & \text{if $\ell'=\ell$}\, ,\\
0 &&& \text{otherwise}\, . 
\end{array}\right.
\label{rest}
\end{equation} 
We can check that the restriction of \eqref{pure} on each fixed point $x_{\ell'}$ is \begin{equation}\label{purelyequiv}
\sum_\ell \pair{\vec w}{\vec n_\ell} c_1(L_\ell)|_{x_{\ell'}} 
=\pair{\vec w}{\vec \epsilon}\, ,
\end{equation} independent of $\ell'$.

Using the equivariant integration formula \eqref{toric-equiv-integral} and the restrictions \eqref{rest}, we can reproduce the well-known intersection numbers
\begin{equation}
D_\ell\cdot D_{\ell'}=\int_{\cM}c_1(L_\ell)c_1(L_{\ell'}) =
\begin{cases}
+1 & \text{if $\ell'=\ell\pm 1$}\, ,\\
 -\pair{ \vec n_{\ell-1}}{\vec n_{\ell+1} } & \text{if $\ell'=\ell$}\, ,\\
 0 & \text{otherwise}\, .
\end{cases}
\end{equation}
We have $c_1(T\cM)=\sum_\ell c_1(L_\ell)=\sum_\ell D_\ell$
and therefore the canonical class is $K=-\sum_\ell D_\ell$.

Let us now consider the compactification of the 6d \Nequals{(2,0)} theory on $\cM$ to have a 2d theory.
As before, to find the anomaly of the 2d theory, we need to integrate $(N^3/24)p_2(R)=(N^3/24) x_1^2 x_2^2$,
where we assume as before that the $\SO(5)_R$ bundle reduces to  $\SO(2)_1\times \SO(2)_2 \subset \SO(5)_R$
and we denote the first Chern classes of $\SO(2)_{1,2}$ by $x_{1,2}$. 
We write
\begin{equation}
 x_a = \Delta_a c_1^{2\rd}(F_R) + c_1(E_a) \, ,
\end{equation}
where $c_1^{2\rd}(F_R)$ is the 1st Chern class of the 2d R-symmetry bundle,
and $E_{a=1,2}$ is a line bundle over $\cM$ specifying the $\SO(5)_R$ flux over $\cM$.
We parameterize $E_a$ by writing them as \begin{equation}
c_1(E_a)= - \sum_\ell \fp^{(\ell)}_a c_1(L_\ell)\, .
\end{equation}

To preserve supersymmetry, we need \begin{equation}
\Delta_1+\Delta_2 = 2, \qquad \fp^{(\ell)}_1+\fp^{(\ell)}_2=1 \quad \forall \ell \, ,
\end{equation}
since the preserved supercharge couples to $x_1+x_2+c_1(T\cM)$, which should equal $2c_1(F_R)$ by definition.
The 2d anomaly polynomial in the large $N$ limit is  then simply \begin{equation}
\cA_{2\rd}\approx \frac{N^3}{24} \int_{\cM} 
(\Delta_1 c_1(F_R) - \sum_\ell \fp^{(\ell)}_1 c_1(L_\ell))^2
(\Delta_2 c_1(F_R) - \sum_\ell \fp^{(\ell)}_2 c_1(L_\ell))^2\, ,
\end{equation}
which can be evaluated using the equivariant integration formula \eqref{toric-equiv-integral} and the restriction \eqref{rest} of $c_1(L_\ell)$ on the fixed points. Explicitly,
\begin{equation}\label{localization}
\cA_{2\rd}\approx \frac{N^3}{24}  \sum_{\ell} \frac{(\Delta_1 c_1(F_R) - \fp^{(\ell)}_1 \epsilon_1^{(\ell)}  - \fp^{(\ell+1)}_1 \epsilon_2^{(\ell)}  )^2
(\Delta_2 c_1(F_R) - \fp^{(\ell)}_2 \epsilon_1^{(\ell)}  - \fp^{(\ell+1)}_2 \epsilon_2^{(\ell)}  )^2}{\epsilon_1^{(\ell)}\epsilon_2^{(\ell)}}  
\, ,
\end{equation}
which, after taking the sum over fixed points, becomes a quadratic polynomial in $c_1(F_R)$ and $\epsilon_{a}=c_1(J_a)$.%
\footnote{As briefly discussed in appendix \ref{sec:gravitationalblocks}, the entropy function based on gravitational blocks introduced in \cite{Hosseini:2019iad}  is the gravitational  counterpart of \eqref{localization}.}

\subsubsection[Examples of compactifications  \texorpdfstring{$(2,0)$}{(2,0)} theory on toric \texorpdfstring{K\"ahler}{Kahler} surfaces]{Example of  compactifications of $(2,0)$ theory on toric K\"ahler surfaces}

As an example of compactifications of the $(2,0)$ theory on toric K\"ahler surfaces, let us consider $\mathbb{F}_{0}=\mathbb{P}^1\times \mathbb{P}^1$. The toric data are
\be
\vec{n}_{1}=(1,0) \, ,\quad\vec{n}_{2}=(0,1) \, ,\quad\vec{n}_{3}=(-1,0) \, ,\quad\vec{n}_{4}=(0,-1) \, .
\ee
%\begin{figure}[h!!!!!!!!!]
%\centering
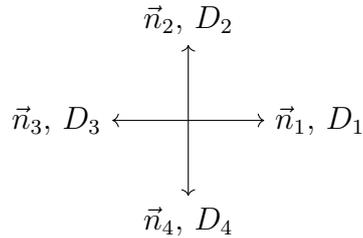
\begin{minipage}[t]{.5\textwidth}
\centering
\vspace{0pt}
\begin{tabular}{|c||c|c|c|c|}
\hline 
\multicolumn{1}{|c}{$\mathbb{F}_{0}$} & \multicolumn{1}{c}{} & \multicolumn{1}{c}{} & \multicolumn{1}{c}{} & \tabularnewline
\hline 
$l$ & 1 & 2 & 3 & 4\tabularnewline
\hline 
$\epsilon_{1}^{(l)}$ & $\epsilon_{1}$ & $\epsilon_{2}$ & $-\epsilon_{1}$ & $-\epsilon_{2}$\tabularnewline
\hline 
$\epsilon_{2}^{(l)}$ & $\epsilon_{2}$ & $-\epsilon_{1}$ & $-\epsilon_{2}$ & $\epsilon_{1}$\tabularnewline
\hline 
\end{tabular}
\end{minipage}\hfill
\begin{minipage}[t]{.5\textwidth}
\centering
\vspace{0pt}
\begin{tikzpicture}
[scale=0.5 ]
\draw[->] (0,0) -- (2,0) node[right] {${\vec n}_1,\, D_1$};
\draw[->] (0,0) -- (0,2) node[above] {${\vec n}_2, \, D_2$};
\draw[->] (0,0) -- (-2,0) node[ left] {${\vec n}_3,\, D_3$};
\draw[->] (0,0) -- (0,-2) node[ below] {${\vec n}_4,\, D_4$};
  \end{tikzpicture}
\end{minipage}
There are only two independent non-equivariant cohomology classes in $H^2(\mathbb{F}_{0})$.
Correspondingly, we define the expansion coefficients as follows:
\be
 \fp_i^{(1)} = \fp_i^{(3)} \equiv \frac{\fs_i}{2} \, , \qquad \fp_i^{(2)} = \fp_i^{(4)} \equiv \frac{\ft_i}{2} \, , \quad \text{ for } i = 1, 2 \, ,
\ee
with
\be
 \fs_1 + \fs_2 = 2 \, , \qquad \ft_1 + \ft_2 = 2 \, .
\ee
The localization formula \eqref{localization} then gives
\bea
 \label{eq:A2d:fromA6d:largeN2II}
  \cA_{2\rd} & \approx \frac{N^3}{12} \left (  \fs_1 \ft_1 \Delta_2^2 + 2( \fs_1\ft_2+\fs_2 \ft_1) \Delta_1 \Delta_2 +  \fs_2 \ft_2 \Delta_1^2\right )  c_1( F_R )^2 \\
 & \qquad + \frac{N^3}{48}    \ft_1 \ft_2 (\fs_1\ft_2+\fs_2 \ft_1)  c_1( J_1 )^2  + \frac{N^3}{48}    \fs_1 \fs_2 (\fs_1\ft_2+\fs_2 \ft_1)  c_1( J_2 )^2  \, ,
\eea 
which correctly reduces to \eqref{eq:A2d:fromA6d:largeN2} for the compactification on $S^2\times \Sigma_\fg$ for $\fg=0$ when we set $c_1( J_2 )=0$. Notice that there is no mixing of the two-dimensional R-symmetry with the rotational isometries. This is due to the fact that, on $\mathbb{P}^1\times \mathbb{P}^1$, the rotational symmetries are enhanced to $\SU(2)\times \SU(2)$.

As a second and less symmetric example, let us consider $\mathbb{F}_{1}$, the blowup of $\mathbb{P}^2$ at a point. The toric data are
\be
\vec{n}_{1}=(1,0) \, ,\quad\vec{n}_{2}=(0,1) \, ,\quad\vec{n}_{3}=(-1,1) \, ,\quad\vec{n}_{4}=(0,-1) \, .
\ee
%\begin{figure}[h!!!!!!!!!]
%\centering
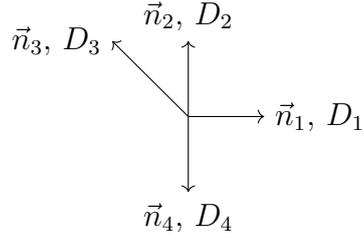
\begin{minipage}[t]{.5\textwidth}
\centering
\vspace{0pt}
\begin{tabular}{|c||c|c|c|c|}
\hline 
\multicolumn{1}{|c}{$\mathbb{F}_{1}$} & \multicolumn{1}{c}{} & \multicolumn{1}{c}{} & \multicolumn{1}{c}{} & \tabularnewline
\hline 
$l$ & 1 & 2 & 3 & 4\tabularnewline
\hline 
$\epsilon_{1}^{(l)}$ & $\epsilon_{1}$ & $\epsilon_{1}+\epsilon_{2}$ & $-\epsilon_{1}$ & $-\epsilon_{2}$\tabularnewline
\hline 
$\epsilon_{2}^{(l)}$ & $\epsilon_{2}$ & $-\epsilon_{1}$ & $-\epsilon_{1}-\epsilon_{2}$ & $\epsilon_{1}$\tabularnewline
\hline 
\end{tabular}
\end{minipage}\hfill
\begin{minipage}[t]{.5\textwidth}
\centering
\vspace{0pt}
\begin{tikzpicture}
[scale=0.5 ]
\draw[->] (0,0) -- (2,0) node[right] {${\vec n}_1,\, D_1$};
\draw[->] (0,0) -- (0,2) node[above] {${\vec n}_2, \, D_2$};
\draw[->] (0,0) -- (-2,2) node[ left] {${\vec n}_3,\, D_3$};
\draw[->] (0,0) -- (0,-2) node[ below] {${\vec n}_4,\, D_4$};
  \end{tikzpicture}
\end{minipage}
We note that the isometry is actually $\SU(2)\times \U(1)$, such that $\epsilon_1$ corresponds to the $\U(1)$ subgroup of $\SU(2)$.

There are again only two independent divisors and thus only two physical fluxes. 
As before, we choose the parameterization
\be
 \fp_i^{(1)} = \fp_i^{(3)} \equiv \frac{\fs_i}{2} \, , \qquad \fp_i^{(2)} = \fp_i^{(4)} \equiv \frac{\ft_i}{2} \, , \quad \text{ for } i = 1, 2 \, ,
\ee
with
\be
 \fs_1 + \fs_2 = 2 \, , \qquad \ft_1 + \ft_2 = 2 \, .
\ee
Then, the localization formula \eqref{localization} yields
\bea
 \cA_{2\rd} & = \frac{N^3}{12} \left( \Delta_2 \fs_1 ( 2 \Delta_1 \ft_2 + \Delta_2 \ft_1 ) + \Delta_1 \fs_2 ( \Delta_1 \ft_2 + 2 \Delta_2 \ft_1 ) \right) c_1( R )^2 \\
 & \qquad + \frac{N^3}{48} \ft_1 \ft_2 ( c_1 ( J_1 ) + 2 c_1 ( J_2 ) ) ( \Delta_1 \ft_2 + \Delta_2 \ft_1 ) c_1 ( R ) \\
 & \qquad + \frac{N^3}{96} ( \fs_2 \ft_1 + \fs_1 \ft_2 ) \left( 2 \fs_1 \fs_2 c_1 ( J_1 )^2 + \ft_1 \ft_2 \left( c_1 ( J_1 )^2 + 2 c_1 ( J_1 ) c_1 ( J_2 ) + 2 c_1 ( J_2 )^2 \right) \right) \, .
\eea
This time we see that the rotational symmetries $J_1$ and $J_2$ mix with the R-symmetry. To find the exact one, we  write the trial central charge
\bea
 c_r(\Delta_i, \epsilon_i)  & = \frac{N^3}{2} \left( \Delta_2 \fs_1 ( 2 \Delta_1 \ft_2 + \Delta_2 \ft_1 ) + \Delta_1 \fs_2 ( \Delta_1 \ft_2 + 2 \Delta_2 \ft_1 ) \right)  \\
 & \qquad + \frac{N^3}{8} \ft_1 \ft_2 ( \epsilon_1  + 2 \epsilon_2 ) ( \Delta_1 \ft_2 + \Delta_2 \ft_1 ) \\
 & \qquad + \frac{N^3}{16} ( \fs_2 \ft_1 + \fs_1 \ft_2 ) \left( 2 \fs_1 \fs_2 \epsilon_1^2 + \ft_1 \ft_2 \left( \epsilon_1^2 + 2 \epsilon_1  \epsilon_2  + 2 \epsilon_2^2 \right) \right) ,
\eea
where we are slightly abusing the notation by using $\epsilon_a$ for the mixing parameter with the trial R-symmetry and the rotational symmetry $J_a$, via $c_1(J_a)=\epsilon_a c_1(F_R)$.
We now extremize the trial central charge with respect to $\Delta_i$ and $\epsilon_a$ under the constraint $\Delta_1 + \Delta_2 = 2$.
The critical points are given by
\bea
 & \bar \Delta_1 = \frac{8 \fs_1^2 (\ft_1-\ft_2) \ft_2+8 \fs_2 \fs_1 \ft_1 (\ft_1-2 \ft_2)+2 \ft_1^2 \left(\ft_2 (\ft_2-\ft_1)-4 \fs_2^2\right)}
 {4 \fs_1^2 \ft_2 (\ft_1-2 \ft_2)+4 \fs_2 \fs_1 \left(\ft_1^2-4 \ft_2 \ft_1+\ft_2^2\right)-\ft_1 \left(\fs_2^2 (8 \ft_1-4 \ft_2)+(\ft_1-\ft_2)^2 \ft_2\right)} \, ,\\
 & \bar \epsilon_1 = 0 \, , \\
 & \bar \epsilon_2 = \frac{8 \left(\fs_2 \ft_1^2+\fs_1 \ft_2^2\right)}{4 \fs_1^2 \ft_2 (\ft_1-2 \ft_2)+4 \fs_2 \fs_1 \left(\ft_1^2-4 \ft_2 \ft_1+\ft_2^2\right)-\ft_1 \left(\fs_2^2 (8 \ft_1-4 \ft_2)+(\ft_1-\ft_2)^2 \ft_2\right)} \, .
\eea
The exact central charge then reads
\be
 c_{\text{CFT}} = 2 N^3 \frac{(\fs_2 \ft_1+\fs_1 \ft_2) \left(4 \fs_2^2 \ft_1^2+4 \fs_1 \fs_2 \ft_2 \ft_1+\ft_2^2 \left(4 \fs_1^2-\ft_1^2\right)\right)}
 {\ft_1 \left(\fs_2^2 (8 \ft_1-4 \ft_2)+(\ft_1-\ft_2)^2 \ft_2\right) - 4 \fs_1^2 \ft_2 (\ft_1-2 \ft_2) - 4 \fs_2 \fs_1 \left(\ft_1^2-4 \ft_2 \ft_1+\ft_2^2\right)} \, .
\ee
We pause here to mention that $\bar \epsilon_1=0$ because the corresponding $\U(1)$ rotational symmetry is part of an $\SU(2)$ isometry, and therefore cannot mix with the R-symmetry.

In the previous examples we chose a representative for the physical fluxes. It is easy to see that  turning on purely equivariant bundles would  not affect 
the result. It follows indeed from \eqref{purelyequiv} that purely equivariant fluxes can be reabsorbed in a redefinition of $\Delta_i$ and, in particular, they
do not affect the value of the exact central charge. 

We should note that positivity of the central charge after $c$-extremization does not guarantee alone that the 2d CFT exists. Examples where $c$-extremization fails are given in \cite{Benini:2013cda}.
In particular, supergravity solutions based on Einstein-K\"ahler surfaces and a single flux along $c_1 (T \cM)$ (all $\fp_i^{(\ell)}$ equal) exist only for surfaces with negative curvature,
thus suggesting the corresponding compactifications on positively curved surfaces are unstable at large $N$, although $c$-extremization gives a positive central charge $c_{\text{CFT}}$ in various cases, including examples for $\mathbb{P}^1\times \mathbb{P}^1$ and $\mathbb{P}^2$.

\section[Rotating black strings in AdS\texorpdfstring{$_5$}{(5)} and AdS\texorpdfstring{$_7$}{(7)}]{Rotating black strings in AdS$_5$ and AdS$_7$}
\label{subsec:BSADS7}

We now switch gear and turn to the bulk duals of some of the field theory results obtained so far.
In particular, we focus on the \emph{rotating} black strings with AdS$_7 \times S^4$ asymptotics, that are holographically dual to the 6d $\cN = (2,0)$ theory on $T^2 \times S^2\ww \times \Sigma_{\fg}$.
These solutions are a two-parameter generalization of the black strings with a topological twist on $T^2 \times \Sigma_{\mathfrak{g}_1} \times \Sigma_{\mathfrak{g}_2}$ in \cite{Benini:2013cda}
that include an extra free electric charge parameter in addition to the angular momentum.%
\footnote{Note that the refinement by angular momentum only exists on $S^2$ and not at higher genera.}
We are in search for an analytic form of the near-horizon geometries of these black strings, and will only argue on general grounds about the existence of a full flow interpolating between the horizon and the asymptotic spacetime that as a rule can be only constructed numerically. From supergravity perspective it then turns out to be convenient to use the dimensional reduction along the Riemann surface down to five dimensions, already performed in \cite{Szepietowski:2012tb}. We can then go one step further and reduce along the length of the black strings to a four-dimensional black hole carrying momentum along the compactification circle, similarly to the path taken in \cite{Hosseini:2018uzp}. This will prove equally useful in the comparison with the microscopic results since the resulting four-dimensional black hole entropy can be directly matched with the {\it charged Cardy formula} we reviewed in section \ref{sec:Cardy}. Note that due to the passage via five dimensions, we would also be able to relate the same near-horizon geometry to AdS$_5$ asymptotics as indicated schematically on the figure below.

\begin{figure}[ht]
\centering	
\begin{tikzpicture}[scale=1, every node/.style={scale=.9}]
	
	\node at (-2.3,0.2){maximal};
	\node at (-2.3,-0.3){7d sugra};
	
	\draw[fill = white,thick, rounded corners] (-1.2,-.75) rectangle (1.5,.75);
	\node at (0.15,0){\Large AdS$_7$};
	
	\draw[->,>=stealth, red, ultra thick] (0.15,-.85) -- (0.15,-2.1);

	\node at (-3,-2.8){STU $+$ UHM \cite{Szepietowski:2012tb}};
	\node at (-3,-3.3){5d ${\cal N} = 2$ sugra};
	
	\node at (0.6,-1.5){\large $\Sigma_{\mathfrak g}$};		

	\draw[fill = white,thick, rounded corners] (-1.2,-3.75) rectangle (1.5,-2.25);
	\node at (0.15,-3){\Large AdS$_5$};
		
	\draw[->,>=stealth, dashed, cadmiumgreen, ultra thick] (1.6,-3) -- (3.6,-3);	

	\draw[->,>=stealth, dashed, blue, ultra thick] (1.6,-0.3) -- (4,-2.1);
				
	\draw[fill = white,thick, rounded corners] (3.7,-3.75) rectangle (6.7,-2.25);
	\node at (5.2,-3){\Large BTZ$ \ltimes S^2\ww$};

	\draw[fill = white,thick, rounded corners] (3.7,-6.35) rectangle (6.7,-4.85);
	\node at (5.2,-5.6){\Large AdS$_2 \ltimes S^2\ww$};

	\node at (-3,-5.4){STU $+$ UHM};
	\node at (-3,-5.9){4d ${\cal N} = 2$ sugra};

	\node at (5.6,-4.3){\large $S^1$};

	\draw[<->,>=stealth, orange, ultra thick] (5.2,-3.8) -- (5.2,-4.8);
		
	\draw[->,>=stealth, red, ultra thick]  (4.7, 0.3) -- (5.5, 0.3);
	\node at (7.85, 0.3){Maldacena-Nu\~{n}ez flow \cite{Maldacena:2000mw}};
	
	\draw[->,>=stealth, dashed, cadmiumgreen, ultra thick] (4.7,-0.3) -- (5.5,-0.3);
	\node at (8.2,-0.3){Rotating black string in AdS$_5$};

	\draw[->,>=stealth, dashed, blue, ultra thick] (4.7,-0.9) -- (5.5,-0.9);
	\node at (8.2,-0.9){Rotating black string in AdS$_7$};

	\draw[<->,>=stealth, orange, ultra thick] (4.7,-1.5) -- (5.5,-1.5);
	\node at (8.05,-1.5){The 4d/5d connection \cite{Andrianopoli:2004im,Gaiotto:2005gf,Behrndt:2005he}};
		
	\end{tikzpicture}
	\caption{Supersymmetric solutions in gauged supergravity and the flows that connect them. The direction of the one-sided arrows indicates the decrease of energy scale along the holographic RG flow.}
	\label{trunc_fig}
\end{figure}
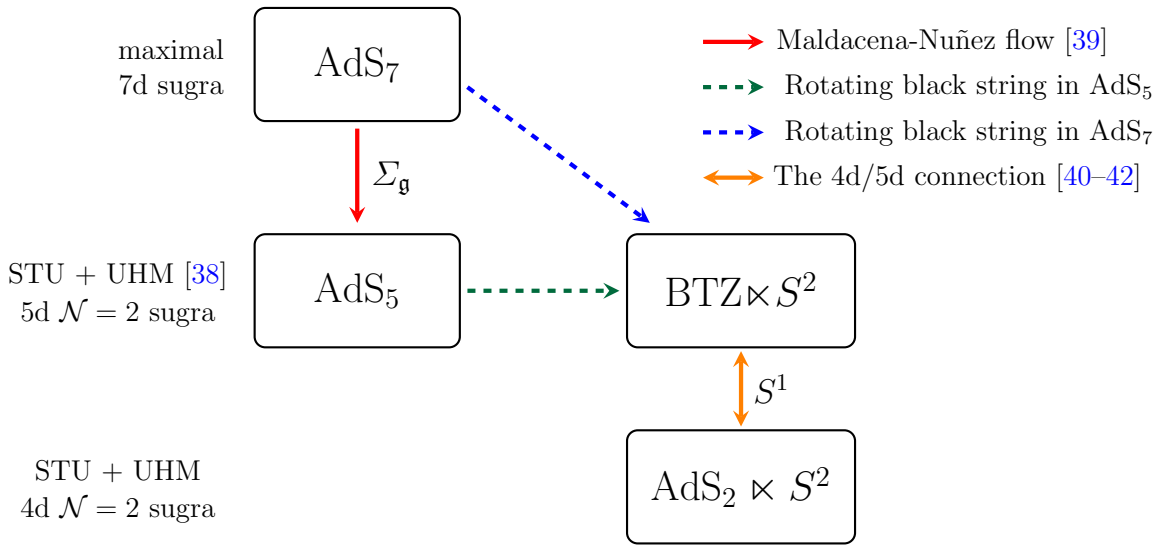

The solutions we are after fall in the general class of rotating black holes with a twist of \cite{Hristov:2018spe}, with one important additional layer of complication.
The resulting four-dimensional supergravity, apart from the three vector multiplets described by the so-called STU model, features one hypermultiplet, called the universal hypermultiplet (UHM).
The non-trivial scalar potential, needed for AdS asymptotics, is realized by gauging two particular abelian isometries of the quaternionic manifold spanned by the hypermultiplet scalars.
This in turn leads to the phenomenon of spontaneous symmetry breaking rendering one of the gauge fields massive, as discussed in \cite{Hristov:2010eu} and \cite{Hosseini:2017fjo}. Similar black hole solutions with charged hypermultiplets were also analyzed in \cite{Halmagyi:2013sla,Chimento:2015rra,Amariti:2016mnz,Guarino:2017pkw,Benini:2017oxt,Bobev:2018uxk}.

Choosing to look for the relevant near-horizon geometry in 4d supergravity makes the problem at hand very similar to the one for rotating AdS$_5 \times S^5$ black string solutions discussed in \cite{Hosseini:2019lkt}.
These are holographically dual to four-dimensional ${\cal N}=4$  SYM on $T^2 \times S^2\ww$ and they can be obtained in a four-dimensional supergravity based on the STU model and no hypermultiplets. The corresponding solutions 
can be obtained from the general setting discussed in this section by setting the hyperscalars to constants and the corresponding killing vectors to zero. We will recover and review these solutions in section \ref{AdS5sol} for future reference.
 
\subsection{The 4d supergravity model}

Before going in details about the supersymmetry preserving Higgs mechanism and the way it allows us to find the solutions, let us first turn to the proper definition of the four-dimensional model we are interested in.
We employ a series of three consecutive dimensional reductions, partially sketched in the figure \ref{trunc_fig},
starting from the Kaluza-Klein reduction of eleven-dimensional supergravity on $S^4$ down to seven-dimensional maximal $\SO(5)$ gauged supergravity \cite{Pilch:1984xy,Pernici:1984xx}.
The next step is the reduction on a Riemann surface $\Sigma_{\mathfrak g}$ of an arbitrary genus $\fg$ down to five-dimensional ${\cal N}=2$ gauged supergravity \cite{Szepietowski:2012tb}, which via the 4d/5d connection \cite{Andrianopoli:2004im,Gaiotto:2005gf,Behrndt:2005he} on a spatial circle we can also view it as four-dimensional ${\cal N}=2$ gauged supergravity. 

The four-dimensional model we are interested in, which very closely resembles its five-dimensional origin, consists of the ${\cal N}=2$ gravity multiplet (bosonic fields: the metric $g_{\mu \nu}$ and a $\U(1)$ gauge field $A^0_\mu$) with additional three vector multiplets (each with a complex scalar $z^i$, $i=1,2,3$,  and a $\U(1)$ gauge field $A^i_\mu$) and one hypermultiplet (with four real scalars $q^u, u =1, .., 4$). We follow the notation and conventions of \cite{Andrianopoli:1996cm}, where one can find all details of the theory with general hypermultiplet gaugings. The quantities in the vector multiplet sector, as well as the supersymmetric solution, are most conveniently written in a symplectic duality-covariant language in terms of vectors. The complex scalars can be parametrized by symplectically covariant sections $X^i/X^0 \equiv z^i$, such that the corresponding scalar manifold is uniquely determined by the so-called prepotential ${\cal F} (X^I)$, $I = \{0, i\}$, of special K\"{a}hler geometry. More specifically, in the case of the STU model we are dealing with, the prepotential is given by
\be
 \label{eq:STUprepotential}
 \cF \left( X^I \right) = \frac{X^1 X^2 X^3}{X^0} \, ,
\ee    
which parametrizes the manifold $[\SU(1,1)/\U(1)]^3$. Furthermore, we need to extensively use the formalism where various vectors, such as the vector of the symplectic section $\{ X^I, F_I \equiv \partial \cF / \partial X^I \}$, or the vector of electromagnetic charges $\Gamma = \{ p^I; q_I \}$, are contracted amongst each other quartically in a symplectically invariant way. For these purposes we need the definition of the quartic invariant $I_4$, which in the case of the STU model and the example of the quartic contraction of the charge vector $\Gamma$, reads%
\footnote{The electromagnetic charge vector $\Gamma = \{ p^I ; q_I \}$ is defined as $\Gamma = \frac{1}{4 \pi} \int_{S^2} F$, where $F$ is the symplectic vector of spatial field strengths.}
\begin{align}
\label{eq:I4electricstu}
\begin{split}
  I_4 (\Gamma) =4\? q_0 p^1 p^2 p^3 - \sum_{i = 1}^3 (p^i q_i)^2  + 2 \sum_{i<j}^3 q_i p^i q_j p^j
- p^0 \bigg( 4\? q_1 q_2 q_3 + p^0 (q_0)^2 + 2\? q_0\? \sum_{i=1}^3 p^i q_i \bigg) . 
\end{split}
\end{align}
For a more pedagogical introduction to symplectic covariance, the symplectic inner product $\Iprod{\cdot}{\cdot}$, the $I_4$-formalism, along with various useful identities, see the appendices of \cite{Bossard:2013oga,Halmagyi:2014qza,Hristov:2018spe,Hosseini:2019iad}. 

The four real scalars that make up the bosonic content of the universal hypermultiplet parametrize the coset space $\SU(2,1)/\U(2)$. In one of the standard conventions that we follow here, the metric on this space can be written in terms of the real coordinates $\{ \phi, \sigma, \zeta, \tilde{\zeta} \}$,
\begin{equation}
\label{eq:UHMmetric}
\rd s_{\textrm{hyp}}^2 =   \rd \phi^2 + \frac{1}{4} \ex^{2\phi}  \left( \rd \zeta^2 + \rd \tilde{\zeta}^2 \right) 
+ \frac{1}{4} \ex^{4 \phi} \left[ \rd \sigma + \tfrac{1}{2}  ( \tilde{\zeta} \rd \zeta - \zeta \rd \tilde{\zeta}  ) \right]^2  \, .
\end{equation}
The isometry group $\SU(2,1)$ has eight generators; two of these are used for gauging in the models we consider explicitly below, generating the group $\mathbb{R} \times \U(1)$. The corresponding Killing vectors read
\begin{align}
	k^{\mathbb{R}} = \partial_{\sigma}\ , \qquad k^{\U(1)} = -\tilde{\zeta} \partial_{\zeta} + \zeta \partial_{\tilde{\zeta}} \, .
	\label{killing}
\end{align}
These two isometries are each gauged by a particular linear combination of the vector fields.
It is then natural to construct the symplectic vectors $\{k^u{^I}; k^u_I \}$, which define the theory uniquely together with the choice of prepotential \eqref{eq:STUprepotential} and the hypermultiplet metric \eqref{eq:UHMmetric}.
For the case at our disposal, the gauging of the two isometries described above is realized explicitly in the following way \cite{Szepietowski:2012tb},
\bea
\label{eq:killingszepietowski}
k_{1} &= - \frac{(\kappa-z_1)}{2 m}\? k^{\mathbb{R}} + m\?  k^{\U(1)} \, , &
k_{2} &= -  \frac{(\kappa+z_1)}{2 m} k^{\mathbb{R}} + m\?  k^{\U(1)} \, ,
\\
k_3 &=  - m \? k^{\mathbb{R}} \, , &
 k_0 &= k^{0, 1, 2, 3} = 0 \, .
\eea
Here $\kappa$ denotes the scalar curvature of the Riemann surface ($\kappa = 1$ for ${\mathfrak g} = 0$, $\kappa = -1$ for ${\mathfrak g} > 1$),
$z_1$ specifies the magnetic flux on $\Sigma_{\mathfrak g}$, and $m$ is the 7d supergravity coupling constant.
The mapping between the parameters here and those in \cite{Szepietowski:2012tb}
are given by
\be
 \label{siepfluxes}
 p_1^{\text{there}} = -\frac{\kappa -z_1}{4m}\, ,\qquad p_2^{\text{there}} = -\frac{\kappa +z_1}{4m} \, ,
\ee
where $p_{1,2}^\text{there}$ are  the magnetic fluxes on $\Sigma_\fg$  for the gauge fields $A^1$ and $A^2$ corresponding to the Cartan generators of $\SO(5)$.
The constraint $p_1^{\text{there}}+p_2^{\text{there}} =-\frac{\kappa}{2m}$ is the twisting
condition imposed by supersymmetry. 
In the other parts of this paper, we use the notation $\fs_{1,2}=|\fg-1|(\kappa\pm z_1)$ instead, see \eqref{ours}.
Note also that the vector field $A^{(0)}_\mu$ in \cite{Szepietowski:2012tb} corresponds to our $A^3_\mu$, while $A^0_\mu$ here is reserved for the Kaluza-Klein vector coming from the extra reduction to four dimensions.

From the quaternionic Killing vectors one can further define a triplet of moment maps $P^x$, $x = 1,2,3$, that are required to write down the supergravity Lagrangian, see \cite{Andrianopoli:1996cm}.
The moment maps associated to the two different gauged isometries \eqref{killing} are given by
\begin{align}
	P^{\mathbb{R}} = \left( 0,  0,   -\frac{1}{2} \ex^{2\phi} \right) , \qquad 
P^{\U(1)} = \left( \tilde{\zeta} \ex^{\phi},  - \zeta \ex^{\phi},   1 - \frac{1}{4} (\zeta^2 + \tilde{\zeta}^2) \ex^{2\phi}\right) , 
\end{align}
and we discuss them in more details in relation to the explicit solution we are after.

\subsection{Near horizon parameters and the BPS Higgsing process}

Let us now focus on the particular supersymmetric solution of the type AdS$_2 \ltimes S^2\ww$, which will be the near-horizon geometry at the end of the black string flow.
For now we assume that the moment maps defined above will only be non-vanishing along the third direction, $P^3 \neq 0$, such that we can solve for the metric, gauge fields and vector multiplet scalars as in \cite{Hosseini:2019lkt}.
We discuss the conditions coming solely from the vector multiplets first,
then go on to the conditions coming from the universal hypermultiplet.

\subsubsection{Supersymmetry conditions coming from the vector multiplets}
We are interested in the following set of electromagnetic charges,
\begin{equation}
 \label{eq:chargevector}
 \Gamma = \{ p^0 = 0,p^i; q_0, q_i \} \, ,
\end{equation}
which, together with the conserved angular momentum $\cJ$, will eventually specify the complete supersymmetric solution.
In terms of the 5d/7d rotating black strings, the $p^i$ and $q_i$ are the magnetic and electric charges, while $q_0$ has the interpretation of momentum added along the string direction.

The four-dimensional metric we are interested in has the form
\be
\rd s^2_4 = -\ex^{2\u} \left(r \? \rd t - \frac{\j }{\v }\? \sin^2 (\theta)\? {\rm d} \varphi  \right)^2 + \ex^{-2\u}\?\left( \frac{\rd r^2}{r^2} + \v ^2 \left(\rd \theta^2 + \sin^2(\theta)\, \rd \varphi^2 \right) \! \right) ,
\ee
where
\be
 \ex^{-2\u} = \sqrt{I_4({\cal I}_0)} \, , \quad \v \?{\cal I}_0 = {\cal H}_0 + \j \? P^3 \cos (\theta) \,,\quad \v  =  \Iprod{P^3}{{\cal H}_0} \,  ,
\ee
such that for now the arbitrary symplectic vector of constant parameters $\cH_0$ and  the extra parameter $\j $ specify completely the metric. In the above formulae we already imposed regularity of the metric and absence of conical singularities near the poles, which imply the constraints 
\be
\label{eq:twistedconstraint}
	\Iprod{\cH_0}{I_4' (P^3)} = \Iprod{P^3}{I_4' (\cH_0)} = 0 \, .
\ee

The symplectic sections at the horizon, after a suitable gauge choice, are given by
\be\label{horsec} 
	 \{X^I; F_I \} = - \frac{1}{2 \sqrt{I_4({\cal I}_0)}}\? I_4'({\cal I}_0) + \ii\? {\cal I}_0 \, .
\ee
Ultimately, the solution is uniquely fixed in terms of the conserved electromagnetic charges $\Gamma$ and the angular momentum $\cJ$ from the attractor equations
\be
\label{eq:attr-twisted}
  \Gamma = \frac{1}{4} I^\prime_4\left({\cal H}_0, {\cal H}_0, P^3 \right) + \frac{1}{2}\? \j ^2\? I^\prime_4\left( P^3 \right) ,
\ee
and
\be
 \label{eq:bigJ-twisted}
 {\cal J} = - \frac{\j }{2} \Big( \Iprod{I_4^\prime(P^3)}{I_4^\prime(\cH_0)} - \frac12 I_4(\cH_0, \cH_0, P^3, P^3) \Iprod{P^3}{\cH_0} \rule[.1cm]{0pt}{\baselineskip} \Big) \, ,
\ee
that can be used to determine the parameter $\j $ and the vector $\cH_0$.
Moreover, we should impose the condition coming from the supersymmetric twist in the case where the horizon has spherical topology,
\be
\label{eq:twistingcondition}
	\Iprod{P^3}{\Gamma} = -1 \, .
\ee 
The allowed conserved charges are therefore constrained by this \emph{twisting condition}, as well as by the constraints \eqref{eq:twistedconstraint} that decrease the parameter space of charges for regular black holes.

\subsubsection{Supersymmetry conditions coming from the universal hypermultiplet}
Now that we have characterized fully the near-horizon in the absence of the universal hypermultiplet, we should see how the additional degrees of freedom are fixed in a supersymmetric way.
The same two isometries of the UHM that are used for gauging in the present model show up in the consistent truncation to four dimensions of massive type IIA theory on $S^6$.
Therefore we can treat them as in \cite{Hosseini:2017fjo}, in particular noticing that the two gauged isometries give rise to two substantially different physical pictures. While the gauging of only the $\U(1)$ isometry would allow one to decouple completely the hypermultiplet sector from the rest, the gauging of the $\mathbb{R}$ isometry leads to spontaneous symmetry breaking and gives mass to one of the gauge fields. The latter process means that one of the vector multiplets combines with the hypermultiplet to make a single {\it massive} vector multiplet and thus in a BPS way decreases the massless degrees of freedom. 

Dealing with the $\U(1)$ isometry first, notice that upon fixing the two hypermultiplet scalars $\zeta$ and $\tilde{\zeta}$ to zero, the Killing vector and moment maps reduce to
\be
	\zeta = \tilde{\zeta} = 0 \quad \Rightarrow \quad k^{\U(1)} = 0\ , \quad P^{\U(1)} = \left( 0, 0, 1 \right) ,
\ee
such that there is no real coupling between the hypermultiplet sector and the vector multiplet sector anymore and one can directly use the formulae on the previous page in order to obtain a supersymmetric solution.

The situation with the $\mathbb{R}$ isometry is clearly different since the Killing vector $k^{\mathbb{R}}$ cannot vanish and thus the scalar $\sigma$ is always coupled under the linear combination $k^\sigma_i A^i_\mu$, giving this particular vector field a mass proportional to the non-vanishing $\ex^{4 \phi}$. The scalar $\sigma$ is then the Goldstone boson that gets eaten up by the new massive vector. Imposing maximal supersymmetry in the hypermultiplet sector, the BPS equations \cite{Hristov:2010eu} essentially tell us that the massive vector multiplet decouples from the rest. We need to have covariant constant hyperscalars, implying as integrability condition the following constraint,
\be
\label{eq:chargeconstraint}
	\nabla_\mu q^u = 0 \quad \Rightarrow \Iprod{k^\sigma}{F_{\mu \nu}} = 0\ , \quad \Iprod{k^\sigma}{\Gamma} = 0 \, .
\ee
Here we have already reduced the original Killing vector \eqref{eq:killingszepietowski} to
\be
\label{eq:killings}
	k^\sigma = \bigg\{0;  0,  -\frac{\kappa-z_1}{2 m} , -\frac{\kappa+z_1}{2 m},  -m \bigg\} \, ,
\ee
and corresponding symplectic vectors for the moment maps
\be
\label{eq:moments}
	P^1 = P^2 = 0 \, ,\qquad P^3 = \bigg\{ 0; 0, m + \frac{\kappa-z_1}{4 m}\, \ex^{2\phi}, m + \frac{\kappa+z_1}{4 m}\, \ex^{2\phi}, \frac{m}{2} \ex^{2\phi} \bigg\} \, ,
\ee
as evaluated at the point $\zeta = \tilde{\zeta} = 0$. We additionally require the hyperscalar $\phi$ to be constant, as well as the final constraint from \cite{Hristov:2010eu},
\be
	k^\sigma_i X^i = 0\ ,
\ee
which needs to be applied on the form of the sections as found in \eqref{horsec}. This therefore reduces the parameter space of solutions, as it leads to a number of constraints on the symplectic vector $\cH_0$ labeling the solutions,
\be
\label{eq:H0constraint}	
	\Iprod{k^\sigma}{\cH_0} = \Iprod{k^\sigma}{I_4'(\cH_0)} = I_4 (k^\sigma, P^3, \cH_0 , \cH_0) =  I_4 (k^\sigma, P^3, P^3 , \cH_0) = 0\ .
\ee
Note that in this case the potential additional constraints $\Iprod{k^\sigma}{P^3} = \Iprod{k^\sigma}{I_4'(P^3)} = 0$ are identities of the model.

We can therefore conclude that the full supersymmetric near-horizon geometry, after employing the BPS-Higgsing mechanism, can be found by imposing the attractor equations \eqref{eq:attr-twisted}-\eqref{eq:bigJ-twisted} along with the constraints \eqref{eq:twistingcondition}, \eqref{eq:chargeconstraint}, and \eqref{eq:H0constraint} on our model specified by \eqref{eq:I4electricstu}, \eqref{eq:chargevector}, \eqref{eq:killings} and \eqref{eq:moments}.

Let us finish with the quantity of main interest here, the Bekenstein-Hawking entropy
\be
	S_{\text{BH}} ( \fp^i , q_0 , q_i , \cJ ) = \frac{A}{4 G^{(4)}_{\text{N}}} = \frac{\pi}{G^{(4)}_{\text{N}}} \sqrt{I_4(\cH_0) - \j ^2}\, ,
\ee
which via the attractor equations \eqref{eq:attr-twisted} and \eqref{eq:bigJ-twisted} is a function of $\Gamma$ and $\cJ$.
It is also useful to define the \emph{real} chemical potential conjugate to the angular momentum $\cJ$ as in \cite{Hristov:2018spe},
\be
\label{eq:twsitedchempotJ}
	w \equiv \frac{\j }{\v  \sqrt{I_4(\cH_0) - \j ^2}} \, .
\ee

\subsection{Explicit solution}\label{AdS7sol}
The final solution that we find is parameterized by one additional twist parameter $z_2$ (on top of $z_1$) fixing the three magnetic charges $p^i$, one free parameter $q$ fixing the electric charges $q_i$, as well as the charges $q_0$ and $\cJ$. We parametrize the electromagnetic charges in the following way,
	\bea\label{chargesAdS7}
		\Gamma = \bigg\{0, - \frac{1+z_2}{2 m}, - \frac{1-z_2}{2 m}, \frac{\kappa - z_1 z_2}{2 m^3}; q_0, - q \left(1 - 2 z_2 + \frac{\kappa}{z_1} \right), - q \left(1 + 2 z_2 - \frac{\kappa}{z_1} \right), 2 m^2 \frac{z_2}{z_1} q \, \! \bigg\} ,
	\eea
which allows us a somewhat shorter presentation of the main quantities of interest.

One can easily convert to the notation we used in section \ref{sec:S2Sigma} by identifying 
\begin{equation}
{\fs_{1,2}} = |\fg-1|(\kappa_{\Sigma_\fg} \pm z_1)\, ,  \qquad
{\ft_{1,2}} =  \kappa_{S^2} \pm z_2 \, ,
\end{equation}
where we furthermore have $\kappa_{\Sigma_\fg} = \kappa$, $\kappa_{S^2} = 1$.
The hyperscalar $\phi$ is fixed by 
\be
	\ex^{2\phi} = - \frac{4 m^2 (\kappa + \kappa z_2^2 - 2 z_1 z_2)}{(3+z_2^2) - 8 \kappa z_1 z_2 + z_1^2 (1+3 z_2^2)} \, ,
\ee
while the vector multiplet scalars can be found from the symplectic vector $\cH_0 = \{ \alpha^0 ; \beta_0 \}$ that can be parametrized as follows
\bea
	& \alpha^0 = \bigg\{0,  a_r + a_l, a_l - a_r, \frac{ a_r z_1 - a_l \kappa}{m^2} \bigg\} \, , \\
	& \beta_0 = \bigg\{ b_0, b \left( - \frac{a_l z_1}{ 2 a_r z_1 - a_l \kappa} + 1 \right), b \left( - \frac{a_l z_1}{ 2 a_r z_1 - a_l \kappa} - 1 \right), \frac{2 m^2 b a_r}{2 a_r z_1 - a_l \kappa}\,  \!\! \bigg\} \, .
\eea

The counting of independent parameters in the vector of four-dimensional electromagnetic charges works as follows. The parameters $p^I$  and $q_I$ associated with massive vectors are not conserved and are fixed by the BPS conditions. There are only three massless vectors. The magnetic charge for $A^0$ is zero by construction  and   one extra magnetic charge is fixed by the twisting condition. As usual with BPS AdS black holes, there is one additional constraint on the electric charges. 

Upon defining the quantities
\bea
	\Pi &= (1+3 z_2^2) z_1^2 - 8 z_1 z_2 \kappa + (3+z_2^2) \, , \\
	\Theta &= (\kappa^2 + z_1^2 - 2 z_1 z_2 \kappa) (z_2 \kappa + z_1 + \kappa - 3 z_1 z_2) (z_2 \kappa + z_1 - \kappa + 3 z_1 z_2) \, ,
\eea
we can write
\be
	a_l = - \frac{\kappa - 3 z_1 z_2}{2 m} \sqrt{\frac{\Pi}{\Theta}} \, , \quad a_r = \frac{ z_1 + z_2 \kappa}{2 m} \sqrt{\frac{\Pi}{\Theta}}\ , \quad b = -\frac{\kappa^2+2 z_1^2 - z_1 z_2 \kappa}{z_1} q \sqrt{\frac{\Pi}{\Theta}} \, .
\ee
The explicit expression of $b_0$ is left out due to its length and lack of insight.
Finally, we can evaluate the parameter $\j $ in terms of the conserved angular momentum via \eqref{eq:bigJ-twisted}. It reads
\be
	\j  = \frac{\Pi^{3/2}}{(z_2^2 - 1) (\kappa - z_1 z_2) \Theta^{1/2}} \cJ \, .
\ee
The entropy, in the canonical normalization $m=2$, is then given by
\be\label{AdS7entropy}
	S_{\text{BH}} ( z_1 , z_2 , q_0 , q , \cJ ) = \frac{\pi}{8 G^{(4)}_{\text{N}}}  \sqrt{\frac{\Pi}{\kappa - 3 z_1 z_2} \left( q_0 - \frac{16\, (\kappa - 3 z_1 z_2)}{z_1^2} q^2 - \frac{64\, \cJ^2}{(z_2^2-1) (\kappa - z_1 z_2)} \right)} \, .
\ee
The holographic central charge $c_{\text{sugra}} (z_1 , z_2)$ can also be computed, {\it \`a la} Brown-Henneaux \cite{Brown1986}, as \cite{Benini:2013cda}
\be
	c_{\text{sugra}} = \frac{3}{128 G^{(4)}_{\text{N}}} \frac{\Pi}{(\kappa - 3 z_1 z_2)}
%	= \frac{\eta_1 \eta_2 N^3}{4} \frac{\Pi}{(\kappa - 3 z_1 z_2)}
	\, .
\ee
Note that regular solutions with a positive central charge exist only in the case when $\kappa = -1$ corresponding to $\mathfrak{g} > 1$, see \cite[Fig.\,5]{Benini:2013cda}. 

\subsection[Recovering rotating black strings in AdS\texorpdfstring{$_5 \times S^5$}{(5) x S**5}]{Recovering rotating black strings in AdS$_5\times S^5$}\label{AdS5sol}

A similar solution for the model with prepotential \eqref{eq:STUprepotential} but without  hypermultiplets was presented in \cite{Hosseini:2019lkt}.
The model describes rotating AdS$_5 \times S^5$ black string solutions that generalize the static ones in \cite{Benini:2013cda}
by including angular momentum and two electric charges. The solutions can be recovered in the general setting of this section  
by setting all the hyperscalars to a constant value, the Killing vectors to zero,
and the  moment maps to $P^3=G$,%
\footnote{We set the four-dimensional gauge coupling constant $g_{(4)}=1$.}
a constant symplectic vectors of  Fayet-Iliopoulos parameters (or FI gauging).

We give some details about  the  solution here, since  we will need them later. The model is specified by the same prepotential
\be
 \label{eq:STUprepotential2}
 \cF \left( X^I \right) = \frac{X^1 X^2 X^3}{X^0} \, ,
\ee  
and FI parameters
\be
 \label{eq:G:AdS5xS5}
 G=\{0;0,1,1,1\} \, .
\ee

The vector of electromagnetic charges is
\begin{equation}
 \label{eq:chargevector2}
 \Gamma = \{ p^0 = 0,p^i; q_0, q_i \} \, ,
\end{equation}
where now all vectors are massless and all the $p^I$ and $q_I$ are conserved charges.
One magnetic charge fixed by the twisting condition
\be\label{eq:STUmagneticconstraint}
p^1 + p^2 + p^3 = -1\, ,
\ee
and one extra electric charge by the BPS conditions 
\be
\label{eq:q3relation}
q_1\? p^1\? (1+2 p^1) + q_2\? p^2\? (1+2 p^2) + q_3\? p^3(1+2p^3)=0\, .
\ee

The entropy reads
\be
	\label{STU:entropydyonic}
	S_{\text{BH}} ( p^i, q_0 , q_i , \cJ ) = \frac{\pi}{G^{(4)}_{\text{N}}}\? \sqrt{\frac{-I_4(\Gamma)- {\cal J}^2}{\Theta^{\text{STU}} } }\, .
\ee
with
\bea
\Theta^{\text{STU}} &=  (p^1)^2 + (p^2)^2 + (p^3)^2 - 2(p^1 p^2 + p^1 p^3 + p^2 p^3)\, , \\
\Pi^{\text{STU}} &= (-p^1+p^2+p^3) (p^1-p^2+p^3) (p^1+p^2-p^3)\, .
\eea
This expression simplifies in the purely magnetic case
\be
	\label{STU:entropy}
	S_{\text{BH}} ( p^i, q_0, \cJ )= \frac{\pi}{G^{(4)}_{\text{N}}}\? \sqrt{\frac{-4\? q_0 p^1 p^2 p^3- {\cal J}^2}{\Theta^{\text{STU}} } }\, \equiv \frac{\pi}{G^{(4)}_{\text{N}}}\? \sqrt{\cW} .
\ee
The holographic central charge reads \cite{Benini:2013cda}
\be\label{csugraN4}
	c_{\text{sugra}} = -\frac{12 \sqrt{2}}{G^{(4)}_{\text{N}}} \frac{ p^1 p^2 p^3}{\Theta^{\text{STU}} }
	\, .
\ee

For completeness, and in order to illustrate a simple point, let us also present the form of the near-horizon BTZ$ \ltimes S^2_w$ geometry of the black string solutions in this case. We find,%
\footnote{In our conventions the five-dimensional and four-dimensional gauge fields are related by $A^i_{(5)} = \sqrt{2} A^i_{(4)}$ and the gauge coupling constants in the corresponding gauged supergravity by $g_{(5)} = \frac{1}{\sqrt{2}} g_{(4)}$. Recall that we set $g_{(4)} = 1$.}
\bea
 \label{eq:BTZxS2e}
 {\rm d} s^2_5 = & \frac{1}{g_{(5)}^2} \frac{( - p^1 p^2 p^3 \Pi)^{2/3}}{\Theta^2}
 \left( - r^2 {\rm d} \tau^2 + \frac{{\rm d} r^2}{r^2} + \frac{\cW \Theta^2}{(p^1 p^2 p^3)^2} \Big({\rm d} y
 + \frac{p^1 p^2 p^3}{ \Theta \sqrt{\cW}} \?r {\rm d} \tau \Big)^2  \right) \\
 & + \frac{1}{g_{(5)}^2} \left(-\frac{p^1 p^2 p^3}{\sqrt{\Pi}}\right)^{2/3}  \left( {\rm d} \theta^2 + \sin^2 \theta\, \Big({\rm d} \varphi - \frac{\cJ}{p^1 p^2 p^3} {\rm d} y\Big)^2  \right) ,
\eea
where we used the coordinate rescaling $\tau = - \frac{\Theta^2}{\Pi \sqrt{\cW}}\? t $ in order to write the metric in the standard BTZ coordinates, and suppressed the $\text{STU}$ superscript. 

The form of the near-horizon metric is simple enough to actually see what happens upon a further reduction down to three dimensions without performing such a reduction in full detail.
The fibration would give rise to an additional three-dimensional gauge field $A_{\cJ}$ which is evidently a pure gauge.
Since the BTZ circle described by the $y$-coordinate is non-contractible, the gauge field carries a (three-dimensional) electric charge proportional to $\cJ$.
Indeed, we know that the Chern-Simons term for the gauge field in three dimensions imposes it to locally be a pure gauge,
and therefore we see that the form of the metric \eqref{eq:BTZxS2e} is the most general one we could expect, see also the discussion above (2.31) in \cite{Ammon:2012wc}.
We should note that a careful reduction to three dimensions generalizing \cite{Karndumri:2013dca},
albeit out of the present scope, would still be interesting as an independent way of deriving the Chern-Simons level $k$ of the gauge field $A_{\cJ}$.
The analogous form of the near-horizon metric and subsequent discussion pertains equally well to the previous subsection concerning the black strings in AdS$_7$,
which in the static case were also reduced to three dimensions in \cite{Karndumri:2015sia}.

\section{Black strings microstates and the charged Cardy formula}\label{sec:microstate}

In this section we compare the gravity results we have obtained for the entropy of  the rotating black strings with the charged Cardy formula and the prediction from the anomaly polynomial. We will start by reviewing the case of AdS$_5\times S^5$, where the comparison was already done at the level of the elliptic genus in \cite{Hosseini:2019lkt}, and then we move on to AdS$_7\times S^4$.  

\subsection[Black strings in AdS\texorpdfstring{$_5 \times S^5$}{(5) x S**5}]{Black strings in AdS$_5\times S^5$}
\label{sect:AdS5xS5:holo}

We consider the twisted compactification of ${\cal N}=4$ SYM on $S^2_{\epsilon}$ in the large $N$ limit. In this limit, the anomaly polynomial was computed in \eqref{eq:4dN=4SYMto2d}, which we reproduce here:
\be\label{A2dN4}
 \cA_{2\rd} \approx - \frac{N^2 }{2} (\Delta_1 \Delta_2 \fs_3 + \Delta_2 \Delta_3 \fs_1 + \Delta_3 \Delta_1 \fs_2 ) c_1(F_R)^2  - \frac{N^2}{8} \fs_1 \fs_2 \fs_3 \? c_1 ( J ) ^2 \, ,
\ee
where the fluxes $\fs_i$ (satisfying $\sum_{i=1}^3 \fs_i=2$) and the chemical potentials $\Delta_i$ (satisfying $\sum_{i=1}^3 \Delta_i=2$) are associated with the Cartan subalgebra $\U(1)^3\subset \SO(6)$ of the R-symmetry. Recall that $\Delta_i$ is conjugate to the generator $Q_i$ that assigns charge $+1$ to the chiral field $\Phi_i$ and zero to the others, in the standard ${\cal N}=1$ description of ${\cal N}=4$ SYM.
Notice that  $J$  does not mix with the R-symmetry, since it is part of the non-abelian rotational symmetry $\SU(2)$. 
For the convenience of the reader we also repeat the formulae for the 2d trial central charge and the level of the rotational symmetry $k$ that can be extracted from \eqref{A2dN4} and we already presented in \eqref{scft:ckN4}
\bea
 %\label{ckN4}
c_r(\Delta_i)& = - 3N^2  (\Delta_1 \Delta_2 \fs_3 + \Delta_2 \Delta_3 \fs_1 + \Delta_3 \Delta_1 \fs_2 ) \, , \\
k &= \frac{N^2}{4} \fs_1 \fs_2 \fs_3 \, ,
\eea
and the exact central charge of the 2d CFT  obtained by extremizing $c_r(\Delta_i)$ 
\be\label{cr2} c_{\text{CFT}} =%= \frac{ 3 R_{{\rm AdS}_3}}{2 G_N^{(3)}} = %\frac{ 3 \ex^{ f(r_0) + 2 g(r_0)} {\rm vol} (\Sigma_\fg) }{2 G_N^{(5)}}  = 
  12 N^2 \frac{ \fs_1 \fs_2 \fs_3}{ \fs_1^2+\fs_2^2+\fs_3^2 - 2 \fs_1\fs_2 - 2 \fs_2 \fs_3 -2 \fs_3\fs_1} \, .
\ee

 The elliptic genus of the 2d CFT can be extracted from the refined topologically twisted index of the four-dimensional conformal field theory,
 which is the partition function on $T^2\times  S^2\ww$ with a topological twist along $S^2\ww$ \cite{Benini:2015noa}.
The high-temperature limit of the refined twisted index for ${\cal N}=4$ SYM was computed in \cite{Hosseini:2019lkt} and, at large $N$,  reads 
\be\label{RTTIN4}
\log Z(\tau , \mu, \Delta_i) = \frac{\ii \pi}{12 \tau} \left( c_r(\Delta_i) - \frac{3 \mu^2}{ \pi^2} k \right) ,
\ee
in agreement with \eqref{Zasymp}.
The density of states $\rho(n, J, Q_i)$ is then obtained by extremizing%
\footnote{From now on we absorb the vacuum energy  in the definition of $e_0=n_l - \frac{c_l}{24}$.}
\be
 \label{IextN=4}
 \cI_{\text{QFT}} (\tau, \mu, \Delta_i) = \log Z(\tau , \mu, \Delta_i) - \ii \pi \sum_{i=1}^3 \Delta_i Q_i - \ii \mu J -2\pi \ii \tau e_0 +\lambda \bigg( \sum_{i=1}^3 \Delta_i - 2 \bigg) ,
\ee
and evaluating it at its critical points
\be
 \log \rho(e_0, J, Q_i) =  \cI_{\text{QFT}} (\tau, \mu, \Delta_i)\big|_{\text{crit.}} \, .
\ee
Here we introduced the Lagrange multiplier $\lambda$ to enforce the constraint on the $\Delta_i$.
  
We now compare these results with the gravity prediction obtained in section \ref{AdS5sol}. The democratic basis in \eqref{A2dN4}  allows an easy comparison to gravity since the massless vector fields $A^i\, ,i=1,2,3$ are  associated with the Cartan subalgebra $\U(1)^3\subset \SO(6)$.
In particular,  \eqref{eq:STUmagneticconstraint} implies $\fs_i=-2 p^i$. We will also need the following relations among 5d and  4d quantities \cite{Hosseini:2019lkt}%
\footnote{In a frame with purely electric gauging $G_i$, $i=1,2,3$, (see \eqref{eq:G:AdS5xS5}), the charges are quantized as $2 G_i p^i \in \bZ$ and $q_i / ( 2 G_\text{N}^{(4)} G_i) \in \bZ$, not summed over $i$.}
\bea
\label{match}
 & G^{(5)}_{\text{N}}=2\pi G^{(4)}_{\text{N}} \, , \qquad  J=\frac{1}{2 G^{(4)}_{\text{N}}} {\cal J} \, , \qquad  e_0=\frac{1}{2 \sqrt{2} G^{(4)}_{\text{N}}} q_0 \, , \\
 & Q_i=\frac{1}{ 2 \sqrt{2} g_{(5)} G^{(4)}_{\text{N}}} q_i \, , \quad \text{ for } \quad i = 1, 2, 3 \, .
\eea
Finally,
we will use the well-known holographic relation for AdS$_5\times S^5$
\be  
\frac{\pi}{2 g^3_{(5)} G^{(5)}_{\text{N}}} = N^2 \, .
\ee

From \eqref{csugraN4} we recover the matching $c_{\text{CFT}}=c_{\text{sugra}}$ obtained in \citep{Benini:2013cda}. It is also immediate to see that the entropy for purely magnetically charged
black holes \eqref{STU:entropy} matches the charged Cardy formula
\be 
S_{\text{BH}} ( p^i, q_0, J )= 2 \pi \sqrt{ \frac{c_{\text{CFT}} }{6}\left (e_0 - \frac{J^2}{2 k}\right)} \, ,
\ee
which follows from extremizing \eqref{IextN=4} for $Q_i=0$, $\forall i=1,2,3$.
Notice that the extremization with respect to $\Delta_i$ just sets the trial central charge $c_r(\Delta_i)$ equal to its exact value $c_{\text{CFT}}$.

The dyonic case requires a little more work  \cite{Hosseini:2019lkt}. This time there are two flavor symmetries in addition to the rotational one and we expect a more complicated charged Cardy formula. We also need to take into account that  the electric charges $q_i$ on the gravity side are constrained by \eqref{eq:q3relation}. This relation can be interpreted as follows: the black hole has charge \emph{zero} with respect to the \emph{exact} R-symmetry of the 2d CFT
\be
 R_0 = \sum_{i=1}^3 \bar \Delta_i Q_i \, ,
\ee
where $\bar \Delta_i$ is the critical point of $c_r(\Delta_i)$. Indeed, it is easy to check that $\sum_{i=1}^3 q_i \bar \Delta_i =0$. 

We can then choose  two independent flavor charges $K_1=Q_1-Q_3$ and  $K_2=Q_2-Q_3$ and write the trial R-symmetry as
\be
 R(\Delta_i) = R_0 +a_1 K_1 + a_2 K_2\, .
\ee
From the trial central charge
\be
 c_r(\Delta_i)=3 \tr \gamma_3 R(\Delta_i)^2 = c_{\text{CFT}} +  3 \fs_2 a_1^2 + 3 \fs_1 a_2^2 +3 (\fs_1+\fs_2-\fs_3) a_1 a_2 \, ,
\ee
we easily extract the 't Hooft anomaly matrix%
\footnote{This can be also easily computed from the multiplicity of fermionic zero-modes as in \cite{Hosseini:2019lkt}.} 
\be
 \cA_{AB} =  \tr \gamma_3 K_A K_B
 = N^2 \begin{pmatrix}
  \fs_2 & 1 - \fs_3 \\  1 - \fs_3 & \fs_1
 \end{pmatrix} \, .
\ee
Recall that in our conventions the level matrix is given by $k_{AB} = - \cA_{AB}$.
Then, an explicit evaluation of \eqref{STU:entropydyonic} gives  \cite{Hosseini:2019lkt}
\be \label{entr}
S_{\text{BH}} ( p^i, q_0, q_i, J )= 2 \pi \sqrt{ \frac{c_{\text{CFT}} }{6}\left (e_0 - \frac 12 \sum_{A,B=1}^2 \wt Q_A \? (k^{-1})_{AB} \? \wt Q_B - \frac{J^2}{2 k}\right ) } \, ,
\ee
where $\wt Q_A= Q_A-Q_3$, $A = 1, 2$, in complete agreement with the charged Cardy formula \eqref{eq:susy:Charged-Cardy}.
Of course, this formula also follows from extremizing \eqref{IextN=4}. It is interesting to observe that \eqref{IextN=4} can be extremized  for an  arbitrary assignment of electric charges $Q_i$, but the extremum is real  only when the charge under  the exact R-symmetry is zero,  which corresponds to a macroscopically large black hole.

Not all values of $\fs_i$ correspond to regular supergravity solutions.  Clearly, the exact central charge $c_{\text{CFT}}$ must be positive. In addition to this, regularity of the metric further constrains the $\fs_i$.
The region in the space of fluxes $\fs_i$ corresponding to regular solution has been analysed in \cite[Fig.\,1]{Benini:2013cda}. One can check that, in this region, $k$ is always positive while the matrix $k_{AB}$ has signature $(1,1)$.%
\footnote{For the flavor levels this was already observed in \cite{Benini:2013cda}.} This means that one of the flavor currents is supported on the supersymmetric side  (the right-moving one). Since the index counts excitations in the non-supersymmetric side, it is perhaps surprising  that a right-moving charge contributes to the entropy as in \eqref{entr}.
The final result is consistent with the modular transformations of the elliptic genus \eqref{egt}, which depends on the not necessarily positive definite matrix $k_{AB}$,
but it would be interesting to give a simple physical interpretation of the contribution of right-moving currents to the Cardy formula.%
\footnote{Notice that something similar happens for the MSW black holes \cite{Maldacena:1997de}. In this case, the appearance of right-moving charges in the Cardy formula is explained in terms of   a non-zero right-moving momentum which is  allowed by the $(0,4)$ algebra (see for example \cite{deBoer:2006vg,Gaiotto:2006wm} for details).} 

\subsection[Black strings in AdS\texorpdfstring{$_7 \times S^4$}{(7) x S**4}]{Black strings in AdS$_7 \times S^4$}
\label{sect:AdS7xS4:holo}
 
 We consider now the twisted compactification of the 6d $A_{N-1}$ $(2,0)$ theory  on $S^2_{\epsilon} \times \Sigma_\fg$ in the large $N$ limit. The anomaly polynomial was computed in 
 \eqref{eq:A2d:fromA6d:largeN2} which we reproduce here:
\bea
 \label{A2d:fromA6d:largeN2}
  \cA_{2\rd}[A_{N-1}] & \approx \frac{N^3}{12} \left (  \fs_1 \ft_1 \Delta_2^2 + 2( \fs_1\ft_2+\fs_2 \ft_1) \Delta_1 \Delta_2 +  \fs_2 \ft_2 \Delta_1^2\right )  c_1( R )^2 \\
 & \qquad  + \frac{N^3}{48}    \ft_1 \ft_2 (\fs_1\ft_2+\fs_2 \ft_1)  c_1( J )^2  \, ,
\eea 
with
\be
 \Delta_1 + \Delta_2 = 2 , \qquad \fs_1 + \fs_2 = 2 -2 \fg \, , \qquad \ft_1 + \ft_2 = 2 \, .
\ee
The fluxes $\fs_i$ and $\ft_i$  and the chemical potentials $\Delta_i$  are conjugate to the charges $Q_i$ corresponding to the Cartan subalgebra $\U(1)^2\subset \SO(5)$ of the R-symmetry. 
As before, the non-abelian rotation group containing  $J$  does not mix with the R-symmetry.
We  reproduce for the convenience of the reader the 2d trial central charge, as a function of the flavor chemical potentials, and the level of the rotational symmetry given in \eqref{scft:ck(2,0)}
\bea\label{ckN4}
c_r(\Delta_i) &= \frac{N^3}{2} \left (  \fs_1 \ft_1 \Delta_2^2 + 2( \fs_1\ft_2+\fs_2 \ft_1) \Delta_1 \Delta_2 +  \fs_2 \ft_2 \Delta_1^2\right ) , \\
k &= - \frac{N^3}{24}  \ft_1 \ft_2 (\fs_1\ft_2+\fs_2 \ft_1)\, ,
\eea
and the exact central charge of the 2d CFT
\be%\label{cr} 
c_{\text{CFT}} =%= \frac{ 3 R_{{\rm AdS}_3}}{2 G_N^{(3)}} = %\frac{ 3 \ex^{ f(r_0) + 2 g(r_0)} {\rm vol} (\Sigma_\fg) }{2 G_N^{(5)}}  = 
  2 N^3 \frac{ \fs_1^2 \ft_2^2 + \fs_1 \fs_2 \ft_1 \ft_2 +\fs_2^2 \ft_1^2}{ \fs_1 (2 \ft_2 -\ft
 _1) + \fs_2 (2 \ft_1 -\ft
 _2)} \, .
\ee

The exact R-symmetry of the 2d CFT is given by
\be R_0= \sum_{i=1}^2 \bar \Delta_i Q_i\, ,\ee
where $\bar \Delta_i$ is the critical point of $c_r(\Delta_i)$. Introducing the flavor charge $K=Q_1-Q_2$ and writing
\be
 R(\Delta_i) = R_0 +a K\, ,
\ee
we can rewrite  the trial central charge as 
\be
 c_r(\Delta_i)=3 \tr \gamma_3 R(\Delta_i)^2 = c_{\text{CFT}} + \frac{N^3}{2}   ( \fs_1 (\ft_1-2 \ft_2) + \fs_2 (\ft_2- 2 \ft_1))  a^2 \, .
\ee
From this expression we can extract the flavor symmetry level
\be
 k_{FF} = - \cA_{FF} = - \tr \gamma_3 K^2 = - \frac{N^3}{6} ( \fs_1 (\ft_1-2 \ft_2) + \fs_2 (\ft_2- 2 \ft_1)) \, .
\ee

We now move to the comparison with the gravity results in section  \ref{AdS7sol}.  To compare the results, we will need to discuss the  dictionary between gravity and field theory quantities.
The  identification is complicated by the presence of one massive gauge field in the bulk. This can be identified with \cite[(26)--(28)]{Szepietowski:2012tb}\footnote{ Recall that, for the sake of comparison, the vector $A^{(0)}$ in \cite{Szepietowski:2012tb} corresponds to our $A^3$.}
\be  \tilde A^3 = A^3  +\frac{\kappa -z_1}{4m} A^1+\frac{\kappa +z_1}{4m} A^2 \, .\ee
We can write  an effective theory for the  massless fields at the horizon by eliminating the massive gauge field $\tilde A^3$ and the hypermultiplet degrees of freedom.
As discussed in more details in appendix \ref{sec:gravitationalblocks}, this can be also done at the level of gauged supergravity. The BPS condition $k^\sigma_I X^I = 0$ allows us to eliminate one of the section 
\be
\label{reduct0}
  X^3 = - \frac{1}{2 m^2} \left( (\kappa - z_1) X^1 + (\kappa + z_1) X^2 \right) ,
\ee
and write an effective  prepotential 
\be
	{\cal F}^* (X^I) = - \frac{X^1 X^2 \left( (\kappa - z_1) X^1 + (\kappa + z_1) X^2 \right)}{2 m^2\, X^0} \, .
\ee
This kind of approach has been already used  successfully in \cite{Hosseini:2017fjo,Benini:2017oxt,Hosseini:2018uzp}. 
We can then identify the massless fields $A^1$ and $A^2$ of the effective theory with the Cartan generators of $\SO(6)$. 
Notice that the elimination of $\tilde A^3$ leads to a redefinition of the corresponding electric charges 
\be\label{effch0}
 \tilde q_1=q_1 - \frac{1}{2 m^2} (\kappa - z_1) q_3 \, ,\qquad \tilde q_2=q_2 - \frac{1}{2 m^2}(\kappa + z_1) q_3 \, .
\ee
With this information we can write  the dictionary between field theory and gravity%
\footnote{In a frame with purely electric gauging $P_i^{\U(1)}$, $i=1,2$, (see \eqref{eq:moments:effective}), the charges are quantized as $2 m p^i \in \bZ$ and $\tilde q_i / ( 2 G_\text{N}^{(4)} m) \in \bZ$, not summed over $i$.
Recall also that we set $m = 2$ in our conventions.}
\bea
 J=\frac{1}{2 G^{(4)}_{\text{N}}} {\cal J} \, , \qquad  e_0=\frac{1}{ G^{(4)}_{\text{N}}} q_0 \, , \qquad  Q_1=\frac{1}{ 4G^{(4)}_{\text{N}}} \tilde q_1 \, , \qquad  Q_2=\frac{1}{ 4G^{(4)}_{\text{N}}} \tilde q_2 \, .
\eea
The magnetic fluxes for $A^1$ and $A^2$ on $\Sigma_\fg$ and $S^2$ are given in \eqref{siepfluxes} and \eqref{chargesAdS7}, respectively, and they can be easily converted to our normalizations
\bea
 \fs_{1,2} =  | \fg - 1 |  (\kappa \pm z_1)\, , \qquad \ft_{1,2} = 1 \pm z_2 \, .
 \label{ours}
\eea   
Finally, the four-dimensional Newton constant is given by%
\footnote{The dimensional reduction from seven to four dimensions is done on a Riemann surface and a circle of volume $4 \pi |\fg - 1|$ and $2\pi$, respectively.
We also note the standard AdS$_7$/CFT$_6$ relation for $A_{N-1}$ theories, $N^3 = 3 \pi^2 / ( 16 G_{\text{N}}^{(7)} )$.}
\bea
 G^{(4)}_{\text{N}} = \frac{3}{128|\fg-1| N^3 } \, .
\eea

As a first check,  we find $c_{\text{CFT}}=c_{\text{sugra}}$, in agreement with \citep{Benini:2013cda}.
We want now to write the entropy of the black strings in terms of the conserved charges $J$, $Q_1$ and $Q_2$.
As we already discussed in section  \ref{AdS7sol}, the black string has  only one independent  electric charge, although there are two massless gauge fields.
The interpretation is similar to the   case of black strings in AdS$_5\times S^5$:  the charge under the exact R-symmetry of the 2d CFT is identically zero.
Indeed, using \eqref{effch0} and  \eqref{chargesAdS7}, we can easily check that 
\bea Q_1 \bar \Delta_1+ Q_2 \bar \Delta_2 = \frac{1}{ 4G^{(4)}_{\text{N}}}\left ( \tilde q_1 \bar \Delta_1+\tilde q_2 \bar \Delta_2\right ) =0 \, , \eea
where $\bar \Delta_i$ is the critical point of $c_r(\Delta_i)$.  Then, we can write \eqref{AdS7entropy} as
\be\label{Cardy6d} 
S_{\text{BH}} ( z_1 , z_2 , q_0 , q , \cJ )= 2 \pi \sqrt{ \frac{c_{\text{CFT}} }{6}\left (e_0 - \frac 12  \frac{(Q_1-Q_2)^2}{ k_{FF}} - \frac{J^2}{2 k}\right ) } \, ,
\ee
in agreement with the charged Cardy formula \eqref{eq:susy:Charged-Cardy}.

One can check that, in the region of fluxes corresponding to regular supergravity solutions \cite[Fig.\,5]{Benini:2013cda}, the levels $k$ and $k_{FF}$ are positive, corresponding 
to holomorphic (left-moving) currents in the CFT.\footnote{For the flavor levels this was already observed in \cite{Benini:2013cda}.}
 
It would be interesting to give an independent derivation of the charged Cardy formula from the refined topologically twisted index of ${\cal N}=2$ SYM in five dimensions \cite{Hosseini:2018uzp},%
\footnote{See also \cite{Crichigno:2018adf}.}
which is supposed to reproduce the elliptic genus of the 2d CFT. 

\section{Discussion and outlook}
\label{sect:discussion}

In this paper we discussed how to derive the general  anomaly polynomial for a theory that is obtained by dimensional reduction on a compactification manifold $\cM_d$, including
background fields for the isometry of  $\cM_d$. We have used the resulting anomaly polynomial to match the holographic prediction for a class of charged and rotating black strings in AdS$_5\times S^5$ found in  \cite{Hosseini:2019lkt},  and a similar class of black strings in AdS$_7\times S^4$, which we newly and explicitly construct in this paper, with the charged Cardy formula \eqref{eq:susy:Charged-Cardy}. There are several questions that are left unanswered by our analysis and we leave for future work.

First of all, for the class of theories considered in this paper, supergravity seems to prefer negatively curved manifolds, thus excluding many interesting examples with internal isometries and also suggesting that various examples where $c$-extremization predicts a positive exact central charge are actually unstable. On the other hand, it is true that most of the existing supergravity solutions are based on Einstein-K\"ahler surfaces and on a simple choice of fluxes. It would be very interesting to enlarge the class of supergravity solutions based on Einstein-K\"ahler surfaces introducing  more complicated choices of fluxes or,
consider solutions corresponding to toric surfaces that do not admit Einstein metric, and compare with the previous results based on  $c$-extremization.
In particular, it would be interesting to see whether there are solutions where the isometry of the internal manifold mixes with the R-symmetry of the 2d CFT and how this is realized in the supergravity solution.

On a different note, it would be interesting to compute the refined topologically twisted index  \cite{Hosseini:2018uzp} of ${\cal N}=2$ SYM in five dimensions\footnote{The theory  decompactifies to the $(2,0)$ theory in six dimensions.}
and reproduce the charged Cardy formula discussed in section \ref{sect:AdS7xS4:holo}, in analogy with what was done for ${\cal N}=4$ SYM in \cite{Hosseini:2019lkt}. Evaluating the five-dimensional refined topologically twisted index at large $N$ is a nontrivial problem.  A proposal for finding the saddle point of the  topologically twisted index, which should capture the entropy in the static case, was discussed  in \cite{Hosseini:2018uzp}. It would be interesting to make it rigorous and generalize it to the rotating case, understanding in the process the role of holomorphic blocks and the relation to the entropy functions introduced in \cite{Hosseini:2019iad} and briefly discussed in appendix \ref{sec:gravitationalblocks}.

Another aspect that we have not touched upon in the present paper is holography beyond large $N$. It is conceptually straightforward to perform the integration of the anomaly polynomials we considered at a finite value for the rank of the gauge group, as explicitly shown in several examples here. The corresponding gravitational calculation requires the knowledge of higher derivative corrections to the relevant supergravity solutions. The question of writing down higher derivative supergravity theories is however still open, with various partial results in different dimensions. The knowledge of the coefficients of the anomaly polynomials has an important role in fixing the relevant higher derivative terms and this has been exemplified for the static magnetically charged black strings in AdS$_7 \times$S$^4$ in \cite{Baggio:2014hua}, where the first subleading corrections were successfully shown to agree on the two sides of the duality. It would be interesting to extend this result to the black strings with electric charges and rotation found here in two-derivative supergravity, corresponding to leading order in $N$.

Let us finish the general discussion by stressing again that the charged Cardy formula \eqref{eq:intro:Charged-Cardy}
and its generalization to multiple $\U(1)$'s appear to have been derived many times over in the literature,
and yet to the best of the authors' knowledge have not been employed in their full potential for two-dimensional $(0,2)$ theories.
The formula was perhaps derived most pedagogically in \cite{Kraus:2006nb} based on the assumption of at least $(0,4)$ supersymmetry.
In the case of a single charge $J$ as in \eqref{eq:intro:Charged-Cardy}, for $(0,4)$ theories one needs to use the $\SU(2)_R$ current with a level $k_{\SU(2)} = \frac{c}{6}$ with $c$ being the 2d central charge.
Probably the most well-known case in which the resulting formula was checked holographically is for the BMPV black hole \cite{Breckenridge:1996is} that is described by the D$1$-D$5$ system,
where the charge $J$ corresponds to angular momentum.
The analogous result holds for other spinning black holes, see e.g.\ \cite{Horowitz:1996ac,Emparan:2006it,Haghighat:2015ega}.
In this regard, the gravitational examples of rotating AdS black strings that we presented here generalize the above results in a genuine $(0,2)$ setting.
We successfully performed a non-trivial test of the charged Cardy formula \eqref{eq:intro:Charged-Cardy} where the level $k$ is independent of the central charge $c$.
In the case of multiple $\U(1)$'s mixing among each other, instead, it turns out that the existing $(0,4)$ holographic examples correspond to the addition of multiple electric charges,%
\footnote{One can of course only distinguish between angular momentum and electric charges from the point of view of the four- or five-dimensional geometry, to which this discussion pertains.}
as in the prototypical example \cite{Maldacena:1997de}.
We also found the analogous answer for the AdS black strings when including electric charges for the flavor symmetries,
after having first determined the exact R-symmetry (a preliminary step that is not needed for the $(0,4)$ examples).

\section*{Acknowledgements}

The authors would like to thank Francesco Benini, Nikolai Bobev, Valentin Reys and Chiara Toldo for useful discussions and comments;
and in particular, Masahito Yamazaki for fruitful discussions and collaboration in the early stages of this work.
SMH and  YT are supported in part by WPI Initiative, MEXT, Japan at IPMU, the University of Tokyo.
The work of SMH is also supported in part by JSPS KAKENHI Grant-in-Aid (Wakate-A), No.17H04837 and JSPS KAKENHI Grant-in-Aid (Early-Career Scientists), No.20K14462.
KH is supported in part by the Bulgarian NSF grants DN08/3, N28/5, and KP-06-N 38/11.
YT is supported in part by JSPS KAKENHI Grant-in-Aid (Wakate-A), No.17H04837 and JSPS KAKENHI Grant-in-Aid (Kiban-S), No.16H06335.
AZ is partially supported by the INFN, the ERC-STG grant 637844-HBQFTNCER, and the MIUR-PRIN contract 2017CC72MK003.

\begin{appendix}

\section{Entropy function from gravitational blocks}\label{sec:gravitationalblocks}

A general entropy function for arbitrary charged and rotating AdS black holes  has been proposed in  \cite{Hosseini:2019iad} in terms of gravitational blocks. In this appendix we will show how it works 
in our examples. 

The entropy function of the rotating AdS$_4$ black holes discussed in section  \ref{subsec:BSADS7} is a simple generalization of that discussed in \cite{Hosseini:2019iad} for theories without hypermultiplets.
Let us summarize the result. The entropy function  can be obtained by gluing gravitational blocks
\be\label{bb}
 \cB (X^I , \omega) \equiv - \frac{\cF(X^I)}{\omega} \, ,
\ee
where $\cF(X^I)$ is the gauged supergravity prepotential, and it is  given by 
\be\label{gluing}
 \cI ( p^I , \chi^I , \omega) \equiv  \frac{\pi}{4 G_{\text{N}}^{(4)}} \left (\sum_{\sigma = 1}^2 \cB \big(X^I_{(\sigma)} , \omega_{(\sigma)} \big) -2 \ii \chi^I q_I - 2 \omega \cJ \right ) .
\ee
Here $\chi^I$ and $\omega$ are the chemical potentials conjugate to electric charge $q_I$ and the angular momentum $\cJ$, respectively.
We also need to use the \emph{$A$-gluing} rule
\bea
\label{Agluing}
  X^{I}_{(1)}  &= \chi^I - \ii \omega \? p^I \, ,& \omega_{(1)} &= +\omega \, , \\
  X^{I}_{(2)}  &= \chi^I + \ii \omega \? p^I \, , & \omega_{(2)} &= - \omega  \, ,
\eea
due to the topological twist on the spherical part of the horizon geometry.
The functional $\cI ( p^I , \chi^I , \omega)$ needs to be extremized with respect to the chemical potentials, subject to one additional constraint.
In the presence of hypermultiplets, the constraint proposed in \cite[(3.13)]{Hosseini:2019iad} becomes
 \be
  P^3_I \chi^I = 2 \, ,
 \ee 
with the further restrictions discussed below.
The attractor mechanism works as follows. The  values of the sections at the south pole (SP) at $\theta = 0$ and the north pole (NP) at $\theta = \pi$ of the sphere are given by
\bea
 X^I_{\text{SP, NP}} &= \frac{\ii}{2} \Big( \bar \chi^I \mp \ii \bar \omega p^I \Big) \, , & I& = 0 , \ldots, 3 \, , \\
 \bar \omega &= - 2 w \, ,
\eea
where $w$ is defined in \eqref{eq:twsitedchempotJ}; $\bar \chi$ and $\bar \omega $ are the critical points of the functional $\cI( p^I , \chi^I , \omega )$.
Moreover,
\be
 S_{\text{BH}} (p^I , q_I , \cJ) = \cI ( p^I , \bar \chi^I , \bar \omega) \, .
\ee

For the rotating black strings in AdS$_5\times S^5$ we use the prepotential given in \eqref{eq:STUprepotential} and we extremize the functional with respect to $\chi^I$ and $\omega$ with the constraint 
\be \sum_{i=1}^3 \chi^i=2 \, .\ee

For the rotating black strings in AdS$_7\times S^4$ the BPS-Higgs mechanism leads to an effective reduction of the independent variables and of the supergravity prepotential \cite{Hosseini:2018uzp}.
The BPS condition,
\be
\label{reduct}
 k^\sigma_I X^I = 0 \quad \Rightarrow \quad X^3 = - \frac{1}{2 m^2} \left( (\kappa - z_1) X^1 + (\kappa + z_1) X^2 \right) ,
\ee
allows us to eliminate $X^3$ (and $\chi^3$) from the entropy functional  \eqref{gluing} and from the constraint $P^3_I \chi^I = 2$.
We obtain in this way an effective supergravity prepotential that parameterizes the remaining massless fields on the horizon 
\be
	{\cal F}^* (X^I) = - \frac{X^1 X^2 \left( (\kappa - z_1) X^1 + (\kappa + z_1) X^2 \right)}{2 m^2\, X^0} \, ,
\ee
which can be used in the gravitational block \eqref{bb}.
Since the symplectic vector $k^\sigma$ is proportional to the symplectic vector of moment maps $P^{\mathbb{R}}$,  the condition $k^\sigma_I X^I = 0$ leads to
the constraint
\be
	P^{\U(1)}_I \chi^I = 2 \, ,
\ee
where we suppressed the index $3$ for the moment map triplet. From the explicit expression of $P^{\U(1)}_I$, setting $\zeta=\tilde\zeta=0$, we find the effective moment map
\be
 \label{eq:moments:effective}
	P^{\U(1)} = \{0, 0, 0; 0, m ,m \} \, ,
\ee
in agreement with \cite[(4.17)]{Hosseini:2018uzp}.  Notice that eliminating $X^3$  (and $\chi^3$) leads to a redefinition of the electric charges in the effective entropy functional
\be\label{effch}
 \left\{q_0, \tilde q_1=q_1 - \frac{1}{2 m^2} (\kappa - z_1) q_3, \tilde q_2=q_2 - \frac{1}{2 m^2}(\kappa + z_1) q_3 \right\} .
\ee

We have checked explicitly that the entropy and scalar sections of the explicit solutions in section  \ref{subsec:BSADS7} precisely agree with the proposed entropy function here.
For the rotating black strings in AdS$_5\times S^5$ this was already done in \cite{Hosseini:2019iad}.

It is easy to see that the gluing of gravitational blocks corresponds to the fixed point formula for the equivariant integral of the anomaly polynomial discussed in section \ref{sect:equiv:int:anomaly}.
For example, in the case of rotating strings in AdS$^5 \times S^5$, by identifying
\be
 \chi^i=\Delta_i\, ,\qquad \chi^0= 2 \ii \frac{\beta}{\pi} \, ,\qquad \omega = \ii \frac{\epsilon}{\pi} \, ,
\ee
where $\epsilon = c_1 (J)$, we can rewrite \eqref{gluing} as
\be
 \label{eq:equiv:int:N=4:SYM}
 \cI ( \fs^I , \Delta^I , \tau , \epsilon) =
 \frac{\ii \pi^2 N^3}{4  \tau} \left [ \frac{ \prod_{i=1}^3\left (\Delta_i-\frac{\epsilon}{2\pi} \fs_i\right )  }{\epsilon}
 -\frac{ \prod_{i=1}^3\left (\Delta_i+\frac{\epsilon}{2\pi} \fs_i\right )  }{\epsilon}  \right ]
 - \ii \pi \sum_{i=1}^3 \Delta_i Q_i - \ii \epsilon J -2\pi \ii \tau n  \, ,
\ee
to be extremized under the constraint $\sum_{i=1}^3 \Delta_i=2$.
The quantity in bracket in \eqref{eq:equiv:int:N=4:SYM} is the analogue of \eqref{localization} in the case where $6\rd$ $(2,0)$ theory on $\cM_4$ is replaced by $\cN = 4$ SYM on $S^2$, and the anomaly polynomial \eqref{A6d:largeN} of the six-dimensional theory by \eqref{eq:4dN=4SYM}.

\end{appendix}

\bibliographystyle{ytphys}

\bibliography{5DRTTI}

\end{document}